\documentclass[%
pra,
twocolumn,
 amsmath,amssymb,
 aps,
]{revtex4-2}

\setcitestyle{numbers,square}
\numberwithin{equation}{section}

\usepackage{graphicx}
\usepackage{dcolumn}
\usepackage{bm}
\usepackage{amsmath, mathtools, xparse}
\usepackage{caption}
\usepackage{subcaption}
\usepackage{booktabs}
\usepackage{xcolor}
\usepackage{fancybox}
\usepackage{hyperref}
\usepackage{listings} 
\usepackage{braket}
\usepackage{dsfont}
\usepackage[toc, title, page]{appendix}

\newcommand{\appropto}{\mathrel{\vcenter{
  \offinterlineskip\halign{\hfil$##$\cr
    \propto\cr\noalign{\kern2pt}\sim\cr\noalign{\kern-2pt}}}}}

\begin{document}

\preprint{APS/123-QED}

\title{Stoquasticity in circuit QED}

\author{A. Ciani}
\affiliation{QuTech, Delft University of Technology, P.O. Box 5046, 2600 GA Delft, The Netherlands}
\author{B.M. Terhal}
 \affiliation{QuTech, Delft University of Technology, P.O. Box 5046, 2600 GA Delft, The Netherlands and JARA Institute for Quantum Information, Forschungszentrum Juelich, D-52425 Juelich, Germany}

\date{\today}

\begin{abstract}
We analyze whether circuit QED Hamiltonians are stoquastic, focusing on systems of coupled flux qubits. We show that scalable sign-problem-free path integral Monte Carlo simulations can typically be performed for such systems. Despite this, we corroborate the recent finding \cite{ozfidan2020} that an effective, non-stoquastic qubit Hamiltonian can emerge in a system of capacitively coupled flux qubits. We find that if the capacitive coupling is sufficiently small, this non-stoquasticity of the effective qubit Hamiltonian can be \emph{avoided} if we perform a canonical transformation prior to projecting onto an effective qubit Hamiltonian. Our results shed light on the power of circuit QED Hamiltonians for the use of quantum adiabatic computation and the subtlety of finding a representation which cures the sign problem in these systems.
\end{abstract}

\maketitle

\tableofcontents
\section{Introduction}
An important subject in quantum computational complexity theory is the study of the computational power of quantum Hamiltonians, in particular the hardness of estimating its ground-state energy. Estimating the ground-state energy of quite general quantum Hamiltonians with sufficiently high accuracy is known to be a hard problem for quantum computers (QMA-hard) \cite{kitaevBook}. 

There is however an important subclass of so-called {\em stoquastic} Hamiltonians, first introduced in Ref. \cite{bravyiTerhalDV}, for which the ground-state problem is believed to be easier: rather than being QMA-complete it is StoqMA-complete \cite{Bessen, Cubitt2017}. The class StoqMA is not very well understood, but it is known that ${\rm NP} \subseteq {\rm StoqMA} \subseteq {\rm QMA}$, suggesting that ground-state energy problem for stoquastic Hamiltonians is at least as hard at the ground-state energy problem for classical Hamiltonians (NP-complete) but not as hard as the problem for quantum Hamiltonians which do have a sign problem (QMA-complete). 

A quantum Hamiltonian $H$ is stoquastic in a certain basis $\mathcal{B}=\{\ket{x} \}$ if its entries are real and its off-diagonal elements are all non-positive, \i.e., $\braket{x|H|y}\le 0$ for $x \neq y$ \cite{bravyiTerhalDV}.
If $H$ is stoquastic, it can easily be shown that the Gibbs matrix $\exp(-\beta H)$ is entrywise non-negative for all $\beta >0$ in the basis $\mathcal{B}$ and the partition function $Z(\beta) = \mathrm{Tr}\exp(-\beta H)$ can be written as a sum of products of non-negative weights. In addition, the ground-state of $H$ has non-negative amplitudes in the basis $\ket{x}$. \par

The term stoquastic was introduced to capture that these systems avoid the sign problem: the estimation of the partition function or the energy expectation value in the Gibbs state are amenable to stochastic Monte Carlo methods. 
Of particular interest is the use of stoquastic Hamiltonians for quantum adiabatic computation.
It has been shown that adiabatic computation using only stoquastic frustration-free Hamiltonians can be efficiently classically simulated \cite{bravyiTerhal}, but there are more general adiabatic stoquastic computations whose output can only be obtained using a sub-exponential, hence inefficient, number of classical queries \cite{hastings2020},\cite{GV:nosign}. 
This shows that even when one avoids a sign problem, the power of stochastic Monte Carlo methods can be limited. On the other hand, such methods provide heuristic, often well-performing, classical simulation strategies. We refer the reader to \cite{AL:AC} for a general review on adiabatic quantum computation including the use of stoquastic Hamiltonians. 

Important physical realizations of quantum adiabatic computation in the form of quantum annealing use inductively coupled flux qubits \cite{harris2009}. These coupled flux qubits, described by the formalism of circuit QED, give rise to the effective transverse field Ising model (TIM) of quantum annealing \cite{kadowakiNishimori, Santoro2427, Johnson2011Quantum}. Since the TIM Hamiltonian is stoquastic and hence amenable to quantum Monte Carlo methods, the power of the quantum annealing method is not well understood \cite{isakov2016, AL:AC}. Some research has been devoted to the use of additional non-stoquastic terms in the quantum annealing schedule \cite{hormozi2017, nishimori2017, albash2019, albash2020}, usually referred to as non-stoquastic catalysts. Again, it should be clear that the use of purely stoquastic Hamiltonians does not preclude a quantum computational advantage, in particular when the computation evolves through states other than the ground state via so-called diabatic quantum annealing \cite{crosson2020prospects}.\par

In this paper we study general Hamiltonians that emerge in circuit QED \cite{voolDevoret, blaisCQED, GU20171} and their stoquasticity. In circuit QED one starts with a Lagrangian, and then one constructs a Hamiltonian, expressed in terms of electrical degrees of freedom, such as fluxes and charges which are by definition conjugate variables. Quantization of such system results in a Hamiltonian which is the electric equivalent of a quantum mechanical system with conjugate variables of momentum and position. Such continuous-variable Hamiltonian is then represented in its low-energy discrete sub-space, using perturbative methods, leading to an effective Hamiltonian which emulates a spin system. 

What we find in this paper is that very general flux qubit Hamiltonians, --and even transmon qubit Hamiltonians from a certain perspective--, can be called stoquastic: their thermal properties are directly simulatable using classical Monte Carlo methods. Curiously, this does not imply that a corresponding low-energy effective qubit Hamiltonian is also stoquastic, even allowing for local basis changes on the qubits. Such example of a non-stoquastic qubit coupler for a pair of capacitively and inductively-coupled flux qubits was first presented in \cite{ozfidan2020}. 

However, we also show that for weak coupling, if we apply a canonical transformation on the continuous-variable Hamiltonian {\em before} projecting down to a qubit space, the resulting qubit Hamiltonian is a transverse field Ising model and thus no longer non-stoquastic. 

These results show that apparent sign problems can be cured by transformations and that the power of such transformations can depend at what level they are applied, \i.e., on the global (continuous-variable) Hamiltonian or on the effective qubit Hamiltonian. Previously, the effect of curing the sign problem by local basis changes for qubit Hamiltonians was studied in \cite{marvianLidar, klassenTerhal, klassenMarvian, ioannou2020, hangleiterEisert}. 
It has been an open question whether a stoquastic high-energy `master' Hamiltonian can have a non-stoquastic effective low-energy Hamiltonian, --obtained using Schrieffer-Wollf perturbation theory--, even allowing for local basis changes in the basis of the low-energy Hamiltonian \footnote{One can easily construct an example of a 3-qubit stoquastic Hamiltonian which has a non-stoquastic two-qubit low-energy effective Hamiltonian {\em without} permitting additional local basis changes.}. 

The capacitively- and inductively coupled flux-qubit Hamiltonians thus seem to provide  new examples of such stoquastic master Hamiltonians. Our aim in this paper is not to prove this with full mathematical rigor however: we caution that even the perturbation theory for a simple anharmonic oscillator can be subtle in its convergence \cite{bender-wu}.

The contents of this paper are as follows. In Sec. \ref{sec::cqedflux} we study the general form of circuit QED Hamiltonians and their Lagrangians showing that they never give rise to a sign problem if we work in the continuous flux basis (details in Appendix \ref{app::pimc_flux}). We also discuss a generally non-stoquastic Hamiltonian based on a nonreciprocal electric circuit for contrast and the role of gauge transformations.
Sec.~\ref{sec::stoqEff} deals with the stoquasticity of effective Hamiltonians for coupled flux qubits. The flux qubit itself is reviewed in Appendix \ref{app::fq}. In Sec. \ref{sec::example} we show that two capacitively coupled flux qubits can be described by an effective non-stoquastic Hamiltonian (we also show when this does not happen in Appendix \ref{app::paritysym}).
 
 In Sec. \ref{app::avoidStoq} we then show how in flux qubit systems with weak capacitive and inductive coupling, we can {\em always} project onto an approximate effective qubit transverse-field Ising Hamiltonian which is stoquastic. Crucial to our derivation is the application of an efficient canonical transformation before obtaining this effective model, and the identification of a suitable qubit basis. This procedure is inequivalent to applying local unitaries to cure non-stoquasticity once an effective model is obtained and it explicitly exploits the structure of the initial Hamiltonian. This result is not in contradiction with that of Ref. \cite{ozfidan2020}, since in \cite{ozfidan2020} the capacitive coupling is not weak, and thus the derivation does not directly apply. 
 
 In Sec. \ref{subsec:mcs} we apply the path integral Monte Carlo method to estimate the thermal energy of two capacitively coupled flux qubits. In particular, for weak capacitive coupling, we perform path integral Monte Carlo (PIMC) simulations both in the original flux basis and using the effective stoquastic Hamiltonians, showing good agreement with direct numerical diagonalization. For strong capacitive coupling only PIMC in the flux basis can be used without suffering from the sign problem. Using this method, the average thermal energy can still be accurately estimated, even if the system is described by a low energy effective Hamiltonian which is non-stoquastic. We finally provide some discussion and perspective in Sec. \ref{sec::conclusions}.


\section{Circuit QED Hamiltonians}\label{sec::cqedflux}

In this section we consider typical Hamiltonians in circuit QED \cite{voolDevoret} and discuss their stoquasticity. We focus on Hamiltonians without time-dependent external driving fields as we are interested in thermal and ground-state properties for quantum adiabatic computing. For Hamiltonians subject to time-dependent driving, one can easily break-time reversal invariance, leading to generally complex Hamiltonians \cite{koch:timereversal} which are thus not stoquastic. In addition, electric circuits which are not included in the discussion here and which may lead to non-stoquastic Hamiltonians are ones where phase-slip junctions are present \cite{moojiNazarov, LeGrimsmo}.


We first consider classical Hamiltonians with $N$ independent degrees of freedom of the following form 
\begin{multline}\label{eq::h_gen1}
H = K(\bm{Q}) + U(\bm{\Phi}) = \frac{1}{2} \bm{Q}^T \bm{C}^{-1} \bm{Q} + U(\bm{\Phi}),
\end{multline}
where $\bm{\Phi} = \begin{bmatrix} 
\Phi_1, \dots, \Phi_N 
\end{bmatrix}^{T}$
is the vector of independent fluxes, $\bm{Q} = \begin{bmatrix} 
Q_1, \dots, Q_N 
\end{bmatrix}^{T}$ is the vector of conjugate charges. This Hamiltonian originates from a Lagrangian of the form
\begin{equation}\label{eq::lagran}
 {\cal L}(\dot{\bm{\Phi}},\bm{\Phi})=\frac{1}{2} \bm{\dot{\Phi}}^T\bm{C}\bm{\dot{\Phi}}-U(\bm{\Phi}).
  \end{equation}
 The Hamiltonian is obtained via a Legendre transform $H=\bm{Q}^T \dot{\bm{\Phi}}-{\cal L}$, with the vector of charges defined as $\bm{Q}=\frac{\partial {\cal L}}{\partial \dot{\bm{\Phi}}}$. By definition, through a Poisson bracket, the classical charges and fluxes are conjugate variables. The definition of $\bm{Q}$ for this Lagrangian gives
\begin{equation}
\bm{Q} = \bm{C} \frac{d \bm{\Phi}}{dt}.
\end{equation} 
The capacitance matrix $\bm{C}$ is a $N \times N$ symmetric positive-definite matrix with diagonal entries and negative off-diagonal entries and it is thus invertible \footnote{Using standard methods, the case when ${\bm C}$ is not invertible can be treated separately, \i.e., the modes with zero energy are eliminated as they have no dynamics}. This implies that $\bm{C}^{-1}$ is a symmetric, entrywise non-negative matrix \cite{bermanPlemmons}. 

When we quantize this electric system, we promote the conjugate variables to quantum operators and they obey the canonical commutation relations by definition:
\begin{equation}
[\hat{\Phi}_k,\hat{Q}_l]= i \hbar \delta_{kl}.
\end{equation}
We note that these conjugate operators $\hat{Q}_k$ and $\hat{\Phi}_l$, like momenta and positions in a quantum-mechanical system, take values in $\mathbb{R}$ (see Subsection \ref{sec:tr} and Appendix \ref{app:transmon} for a discussion on the common switch to $2\pi$-periodic phase variables).

The term $ \frac{1}{2} \bm{Q}^T \bm{C}^{-1} \bm{Q}$ in Eq.~\eqref{eq::h_gen1} represents the electrostatic energy stored in the capacitors of the system and, in a mechanical analogy, it has the interpretation of a kinetic term. The term $U(\bm{\Phi})$ represents the inductive, `potential', contribution to the energy. We assume no particular form of $U(\bm{\Phi})$ as our discussion will be general. In circuit QED $U(\bm{\Phi})$ will be given by the sum of the inductive energies of linear self- and mutual inductances, and by the contributions of the (nonlinear) Josephson junctions. We refer the reader to Refs.~\cite{voolDevoret}, \cite{bkd} for a detailed description of how to obtain the Hamiltonian of a superconducting circuit and how this Hamiltonian can be formally written as in Eq.~\eqref{eq::h_gen1}. 

If the inverse capacitance matrix $\bm{C}^{-1}$ is a diagonal matrix, Hamiltonians of the form of Eq. \eqref{eq::h_gen1} are clearly stoquastic in the flux basis when we consider the discretized version of the Hamiltonian operator, i.e. we discretize the flux basis. This can be seen from the fact that the term $U(\bm{\hat{\Phi}})$ is diagonal in this basis, while the kinetic term gives rise to terms $\hat{Q}_k^2=-\frac{\partial^{2}}{\partial \Phi_k^2}$. Discretizing the flux basis so that $\Phi_k=\delta m$ with integer $m$ in some interval, the finite-difference second derivative is approximated as $-\frac{d^{2}f}{d x^2}|\approx \frac{1}{\delta^2}(-f(x+\delta)-f(x-\delta)+2 f(x))$. Hence, as a matrix, this finite-difference negative Laplace operator is real and has non-positive off-diagonal entries. This fact is explicitly used in Ref. \cite{halversonHen2020} to construct a path integral Monte Carlo method. 
When $\bm{C}$ is not diagonal, the discretization becomes more awkward, but, as we show in the next Subsection, we still obtain a sign-problem-free representation of the partition function.

\subsection{Time-Reversal Invariance and Stoquasticity}
\label{sec:t-reversal}

Rather than using a discretization of of the flux degrees of freedom we determine a non-negative path integral expression for $\exp(-\beta H)$ and $Z={\rm Tr} \exp(-\beta H)$ as an integral over non-negative weights for the Hamiltonians in Eq.~\eqref{eq::h_gen1}.

 This representation can be directly used to perform path integral simulations of a quantum adiabatic computation using Hamiltonians of this form. We use this representation in Sec.~\ref{sec::example} to simulate the thermal state of a system of two capacitively coupled flux qubits.
  
Indeed, for the quantum Hamiltonian in Eq.~(\ref{eq::h_gen1}) we can write  $\exp(-\beta H)$ in the flux basis using a Feynman path integral. It can be obtained in discretized form, using Trotterization, see Appendix \ref{app::pimc_flux}, leading to
\begin{multline}\label{eq::pi-stand}
    Z =\int d{\bm{\Phi}} \bra{\bm{\Phi}} e^{-\beta H}\ket{\bm{\Phi}} \approx \\ C \int d\bm{\Phi}_1 \ldots d\bm{\Phi}_M e^{-\beta {\cal H}_c(\bm{\Phi}_1, \ldots, \bm{\Phi}_m)}
\end{multline}
with non-negative constant $C$, periodic boundary conditions $\bm{\Phi}_{M+1} = \bm{\Phi}_{1}$ and classical Hamiltonian
\begin{multline}\label{eq::h_c}
\mathcal{H}_c = \frac{\kappa}{2} \sum_{s=1}^M (\bm{\Phi}_{s+1}^T - \bm{\Phi}_{s}^T)\bm{C} (\bm{\Phi}_{s+1} - \bm{\Phi}_{s})+ \\ \frac{1}{M} \sum_{s=1}^M U(\bm{\Phi}_s),
\end{multline}
with coupling coefficient $\kappa = \frac{M}{\hbar^2 \beta^2}$ assuming large Trotter parameter $M \gg 1$. This expresses the well-known mapping from the partition function of a quantum Hamiltonian with its $N$-dimensional phase space onto the partition function of a $N+1$-dimensional classical Hamiltonian \cite{landauBinder}. 
Taking a continuum limit we can introduce the variable $\tau$, taking values $\tau=\frac{s \beta}{M}$, with $s=1, \ldots, M$ and for large enough number of Trotter slices $M$, the integrand on the r.h.s in Eq.~\eqref{eq::pi-stand} equals 
\begin{equation}
\exp\left(-\int_0^{\beta} d\tau \left[\frac{1}{2\hbar^2}\frac{\partial \bm{\Phi}^T}{\partial \tau}{\bm{C}} \frac{\partial \bm{\Phi}}{\partial \tau}+ U(\bm{\Phi})\right]\right).
\end{equation} 
In this continuum limit we also have
\begin{eqnarray}\label{eq::fm2}
     \bra{\bm{\Phi}_1} \exp(-\beta H ) \ket{\bm{\Phi}_0}= C\int_{\bm{\Phi}_0 \rightarrow \bm{\Phi}_1} D\bm{\Phi} \times \nonumber \\
     \exp\left(\int_0^{\beta}d\tau {\cal L}\left(\frac{i}{\hbar}\frac{d \bm{\Phi}}{d\tau}, \bm{\Phi})\right)\right).
\end{eqnarray}
with Lagrangian ${\cal L}(\dot{\bm{\Phi}}, \bm{\Phi})$.  Clearly, if the integrand ${\cal L}$ on the r.h.s of Eq. \eqref{eq::fm2} is real-valued for all $\bm{\Phi}$, then the r.h.s. is a path integral over non-negative weights and suffers no sign problem. For the time-reversal invariant Lagrangian of Eq.~\eqref{eq::lagran} the integrand is real-valued (as can also be seen from the finite Trotter parameter expressions in  Eq.~\eqref{eq::pi-stand}-\eqref{eq::h_c}). For more general Lagrangians of the form ${\cal L}=K(\bm{\dot{\Phi}})-U(\bm{\Phi})$, --assuming that they lead to a well-defined Hamiltonian--, the path integral seems less useful as the integral over momenta (as in Appendix \ref{app::pimc_flux}) cannot necessarily be executed \cite{kleinert:path-int}. In such cases one could discretize the flux basis and express $\hat{Q}_k=-i \hbar \frac{\partial}{\partial \Phi_k}$ as a finite-difference operator.
In case $H$ can be expanded as a Taylor series in $\hat{Q}$, the Hamiltonian will be real when it only contains terms $\hat{Q}^{2n}$ with $n \in \mathbb{N}$, \i.e., only containing terms which are invariant under time-reversal of operators $\hat{Q}_k \rightarrow -\hat{Q}_k, \hat{\Phi}_k \rightarrow \hat{\Phi}_k$. However, this does not seem sufficient to let $\bra{\bm{\Phi}_1} I-\frac{\beta K(\bm{Q})}{M} \ket{\bm{\Phi}_0}$ be non-negative as the finite-difference expression of, say, a fourth-derivative $Q_k^4$ has alternating signs on the off-diagonal (An example is the Lagrangian of a relativistic, but non-causal, particle, expressed in circuit QED coordinates as ${\cal L}=-mc^2\sqrt{1-\frac{\dot{\Phi}^2}{c^2}}-U(\Phi)$ with Hamiltonian $H=c\sqrt{m^2 c^2+Q^2}+U(\Phi)$).

. \par
  
In circuit QED one also encounters Hamiltonians such as 
\begin{equation}\label{eq::shift-H}
    H_{\rm shift}=\frac{1}{2}(\bm{Q}-\bm{Q}_g)^T \bm{C}^{-1}(\bm{Q}-\bm{Q}_g)+U(\bm{\Phi}).
\end{equation}
where $\bm{Q}_g$ is a vector of classical (gate) charges. This Hamiltonian originates from a Lagrangian of the form
\begin{equation}\label{eq::lagrang-shift}
 {\cal L}_{\rm shift}=\frac{1}{2} \bm{\dot{\Phi}}^T\bm{C}\bm{\dot{\Phi}}+ \bm{Q}_g^T \bm{\dot{\Phi}}-U(\bm{\Phi})
  \end{equation}
 using the definition $\bm{Q}=\frac{\partial {\cal L}_{\rm shift}}{\partial \dot{\bm{\Phi}}}$ and the Legendre transform $H_{\rm shift}=\bm{Q}^T \dot{\bm{\Phi}}-{\cal L}_{\rm shift}$.
 
  It is clear that the Lagrangian ${\cal L}_{\rm shift}$ is not time-reversal invariant due the presence of the charge vector $\bm{Q}_g$.
  When we use $\bm{Q}=-i \hbar \frac{\partial}{\partial \bm{\Phi}}$, the Hamiltonian is complex. 
  
  It is also apparent that a canonical transformation $\bm{Q}'=\bm{Q}-\bm{Q}_g$, $\bm{\Phi}'=\bm{\Phi}$ can bring this Hamiltonian to the form in Eq.~\eqref{eq::h_gen1}. At a quantum level, this transformation preserves the commutation relations between $\bm{\hat{Q}}$ and $\bm{\hat{\Phi}}$ and corresponds to the basis change 
  \begin{equation}\label{eq::bs}
 \ket{\bm{\Phi}'}=e^{i \bm{\Phi}^T \bm{Q}_g/\hbar} \ket{\bm{\Phi}}.    
  \end{equation} Note that then $e^{i \bm{\epsilon}^T \bm{Q}'/\hbar}\ket{\bm{\Phi}'}=\ket{\bm{\Phi}'+\bm{\epsilon}}$ for some vector $\bm{\epsilon}$, as is expected.
  Since the basis change merely applies overall phases, one can verify, following the analysis in Appendix \ref{app::pimc_flux} for $H_{\rm shift}$, that $\bra{\bm{\Phi}} \exp(-\beta H_{\rm shift}) \ket{\bm{\Phi}}$, {\em \i.e., using the original basis}, still has a Monte-Carlo path integral representation with non-negative weights as we start and begin at the same state $\ket{\bm{\Phi}}$. 
  
  However, if we use the original basis $\ket{\bm{\Phi}}$ then $\bra{\bm{\Phi}_1} \exp(-\beta H_{\rm shift}) \ket{\bm{\Phi}_0}$ for $\bm{\Phi}_1 \neq \bm{\Phi}_0$ is complex, and hence $H_{\rm shift}$ cannot be called stoquastic in this basis. The rather trivial basis change to $\ket{\bm{\Phi}'}$ in Eq.~\eqref{eq::bs} cures this, but since the path integral expression for $Z$ uses the same initial and final state one could also omit it. In any case, it follows that for the Hamiltonian in Eq.~\eqref{eq::shift-H} one can apply the path integral Monte Carlo method without sign problem to study the thermal expectation value of $H$ and any diagonal operator in $\bm{\Phi}$.  
      
      \subsubsection{Transmon Qubit}
      \label{sec:tr}
      
   A Cooper-pair box or transmon qubit coupled to an external voltage source, inducing an offset charge $Q_g$, provides a simple example of the Hamiltonian in Eq.~\eqref{eq::shift-H} \cite{koch2007}. In that case we have a single flux $\Phi$ and its conjugate charge $Q$. By the basis change in the previous paragraph the transmon qubit Hamiltonian is thus stoquastic and its thermal state a non-negative matrix (assuming discretization). However, the transmon qubit Hamiltonian is often stated in a rotor subspace of the oscillator space which is spanned by a compact $2\pi$-periodic superconducting {\em phase} basis. This rotor subspace is fixed by the operator $S_Q=\exp(i \pi \hat{Q}/e)$ taking a certain phase eigenvalue, see a detailed analysis in Appendix \ref{app:transmon}. Physical processes which affect the support of the quantum state in these rotor subspaces are the tunneling of single or fractional electron charges through the Josephson junction: these are energetically suppressed due to superconductivity. Even though changes in the support are energetically suppressed, an initial state of a transmon qubit device could well be one with support in multiple rotor subspaces. 
      The upshot of these considerations is this. Whether the transmon qubit can be called stoquastic or not depends on whether one considers the Hamiltonian in a rotor subspace or the full oscillator space and whether one is physically interested in the thermal state in the full oscillator space or the thermal state in a single rotor subspace. In all but one rotor subspace the Hamiltonian is not stoquastic with respect to the phase basis in this subspace and the ground-state wave function is {\em not} a non-negative function of phase.
      
   The flux-type qubits used in quantum annealing \cite{harris2010, harris2009}, \i.e., the focus of this paper, include self- and mutual inductances which makes a switch to a rotor subspace not correct as the dynamics induced by the Hamiltonian is not confined to such subspace.\par
 
\subsubsection{Non-time reversal invariant Hamiltonians}

As another class of examples, we consider the Lagrangian of a so-called non-reciprocal electric circuit which involve gyrators or circulators \cite{rymarz:msc, RBCDV, parra2019}. Such elements can be obtained through active driving \cite{sliwa:circ} or coupling to a magnetic field \cite{VD:gyr}.
The Lagrangian is then of general form
\begin{equation}\label{eq::lgyr}
\mathcal{L}_{\rm gyr}=\frac{1}{2}\dot{\bm{\Phi}}^T \bm{C} \dot{\bm{\Phi}}+\dot{\bm{\Phi}}^T \bm{M} \bm{\Phi}-U(\bm{\Phi}),
\end{equation}
with real, anti-symmetric matrix $\bm{M}$ \cite{rymarz:msc}. Applying a Legendre transformation, one obtains the Hamiltonian
\begin{equation}\label{eq::hgyr}
 H_{\rm gyr}=\frac{1}{2}(\bm{Q}-\bm{M} \bm{\Phi})^T \bm{C}^{-1} (\bm{Q}-\bm{M} \bm{\Phi})+U(\bm{\Phi}).
\end{equation}
It is clear that the Lagrangian $\mathcal{L}_{\rm gyr}$ is not time-reversal invariant due to the term $\dot{\bm{\Phi}}^T \bm{M} \bm{\Phi}$, making the Hamiltonian complex and hence not stoquastic. Following the path integral analysis in Appendix \ref{app::pimc_flux} for $\exp(-\beta H_{\rm gyr})$, one finds complex expressions due to the presence of $\bm{Q}^T \bm{\Phi}$ terms. 

 In this general case the application of canonical (symplectic) transformations, --possibly mixing `positions and momenta' but preserving their commutations relations--, cannot even bring $H_{\rm gyr}$ to a real, time-reversal invariant form, that is, a form in which it is invariant under $\bf{Q}' \rightarrow -{\bf Q}'$. Thus the sign problem for such non-reciprocal Hamiltonians can generically not be cured by a canonical transformation.
 
 \subsubsection{Gauge transformations}
 
 Here we also like to comment on the well-known fact that the Lagrangian ${\cal L}$ does not determine the (quantum) dynamics uniquely, i.e. one can always add a total time derivative of an arbitrary function to the Lagrangian: ${\cal L}(\dot{{\bf \Phi}},{\bf \Phi},t) \rightarrow {\cal L}(\dot{{\bf \Phi}},{\bf \Phi},t)+\frac{df({\bf \Phi},t)}{dt}$. Namely, assuming (for simplicity) that $f$ (nor ${\cal L}$) has no explicit time-dependence, we have $\frac{df({\bf \Phi})}{dt}=\sum_{k=1}^N \frac{\partial f}{\partial {\bm \Phi}_k}\dot{{\bf  \Phi}}_k$. This implies that the conjugate variables ${\bf Q}_k$ get changed to ${\bf Q}'={\bf Q}+\frac{\partial f}{\partial {\bf \Phi}}$ and the Hamiltonian is invariant under the transformation: $H({\bf Q}, {\bf \Phi})=H'({\bf Q}', {\bf \Phi})$ where $H'$ is a different function. 
 
 Thus the gauge freedom expressed in $f$ can lead to a Hamiltonian which seems (at first sight) non-stoquastic. An example is the case $k=1$ with ${\cal L}(\Phi, \dot{\Phi})=\frac{C}{2}\dot{\Phi}^2-U(\Phi)$ and let's take, say, $f(\Phi)=\Phi^3$. We get $H'(Q',\Phi)=\frac{1}{2C}(Q'-3 \Phi^2)^2+U(\Phi)$ which is not manifestly stoquastic. Nonetheless, this still does not lead to a sign problem as Eq. \eqref{eq::fm2} is still satisfied with the new Lagrangian.
 

One can view gauge transformations as potential curing transformations. In fact we can observe that for a Hamiltonian of the form
\begin{equation}
    H=\frac{1}{2}(\bm{Q}-\bm{A}(\bm{\Phi}))^T \bm{C}^{-1} (\bm{Q}-\bm{A}(\bm{\Phi}))+U(\bm{\Phi}).
\end{equation}
where ${\bf A}({\bm \Phi})$ is a $N$-dimensional vector field depending on ${\bf \Phi}$, we can gauge away this field when ${\bf A}=\nabla f({\bf \Phi})=\frac{\partial f}{\partial {\bf \Phi}}$ for some $f({\bf \Phi})$. 

Considering the non-time reversal invariant Lagrangian of the previous section, taking $f({\bf \Phi})={\bf \Phi}^T {\bf M} {\bf \Phi}$ would indeed lead to $\frac{\partial f}{\partial {\bf \Phi}}={\bf M} {\bf \Phi}$, but the anti-symmetry of the matrix ${\bm M}$ immediately implies that $f({\bf \Phi})={\bf \Phi}^T {\bf M} {\bf \Phi}=0$. Said differently, we cannot gauge away these time-reversal symmetry breaking terms, similar as one cannot gauge away the vector potential $\vec{A}=\vec{\nabla} \times \vec{B}$ in a minimal coupling Hamiltonian of a particle in a magnetic field.

The effect of gauge transformations on the form of effective Rabi model Hamiltonians has been recently discussed for instance in \cite{deBernardis2018, distefano2019} and references therein.
Our analysis of how canonical transformations on the full circuit QED Hamiltonian can affect stoquasticity of the effective qubit Hamiltonian bears some resemblance to this discussion, although the focus is different. 
 
\section{Stoquasticity of effective flux qubit Hamiltonians}\label{sec::stoqEff}

In this section we prefer to work with dimensionless variables and we thus introduce dimensionless charges $\bm{q}=\frac{\bm{Q}}{2 e}$, and fluxes $\bm{\phi}=\frac{2 \pi \bm{\Phi}}{\Phi_0}$, with $\Phi_0 = \frac{h}{2 e}$ the superconducting flux quantum. Then, for $k, l=\{1, \dots, N \}$ we then have
\begin{equation}
[\hat{\phi}_k, \hat{\phi}_l] = [\hat{q}_k, \hat{q}_l] =0, \quad [\hat{\phi}_k, \hat{q}_l] = i  \delta_{k l},
\end{equation}
The quantum Hamiltonian of Eq.~\eqref{eq::h_gen1} in terms of the rescaled operators equals 
\begin{equation}\label{eq::h_gen}
H = 4 \bm{\hat{q}}^T \bm{E_C} \bm{\hat{q}}+U(\boldsymbol{\hat{\phi}}).
\end{equation}
where we defined the charging energy matrix 

\begin{equation}\label{eq:charg_en}
\bm{E_C} = \frac{e^2}{2} \bm{C}^{-1}.
\end{equation}

In the previous section we have shown that a general circuit QED Hamiltonian of the form in Eq. \eqref{eq::h_gen} is stoquastic and free of the sign problem and can be simulated by the PIMC algorithm. We note that the Hamiltonian in Eq.~\eqref{eq::h_gen} also models various other `modern' flux qubits such as the fluxonium \cite{Nguyen_2019}.

For superconducting circuits one usually wants to represent the problem using an effective qubit Hamiltonian that describes the behaviour of a discrete number of low-lying energy levels. It is thus natural to ask the question of whether these effective Hamiltonians on qubits are stoquastic or not. Weak inductive coupling in flux qubits gives rise to the Hamiltonian of an effective TIM, \i.e., with $X$, $Z$ and $ZZ$ interactions (see Appendix \ref{app::fq}), which is stoquastic (by applying Pauli $Z$ basis changes so the $X$ terms are negative).

It was shown in \cite{ozfidan2020} that by adding a capacitive coupling between flux qubits, the effective two-qubit Hamiltonian is non-stoquastic and the non-stoquasticity cannot be cured by local unitaries, according to the criteria of Ref. \cite{klassenTerhal}. This finding was further confirmed in Ref. \cite{Consani_2020}, where the authors put forward a more refined analysis based on the perturbative Schrieffer-Wolff (SW) transformation \cite{bravyiDiVincenzo}.

While a higher-order SW transformation is usually necessary in order to achieve good accuracy of all the parameters in the problem, for the purpose of studying stoquasticity, and build intuition, we start by considering effective qubit Hamiltonians that are obtained by simply projecting the initial Hamiltonian onto the computational subspace. This is the SW transformation at lowest-order and it is the common way to obtain an effective qubit model in systems of flux qubits \cite{michielsen2020}. 

We review the basics of flux qubits in Appendix \ref{app::fq}. In the next section, we explain why and under which conditions, a system of two coupled flux qubits can give rise to an effective non-stoquastic Hamiltonian.

\subsection{Two coupled flux qubits with a non-stoquastic effective qubit Hamiltonian} \label{sec::example}

\begin{figure}
\centering
\includegraphics[width=0.48\textwidth]{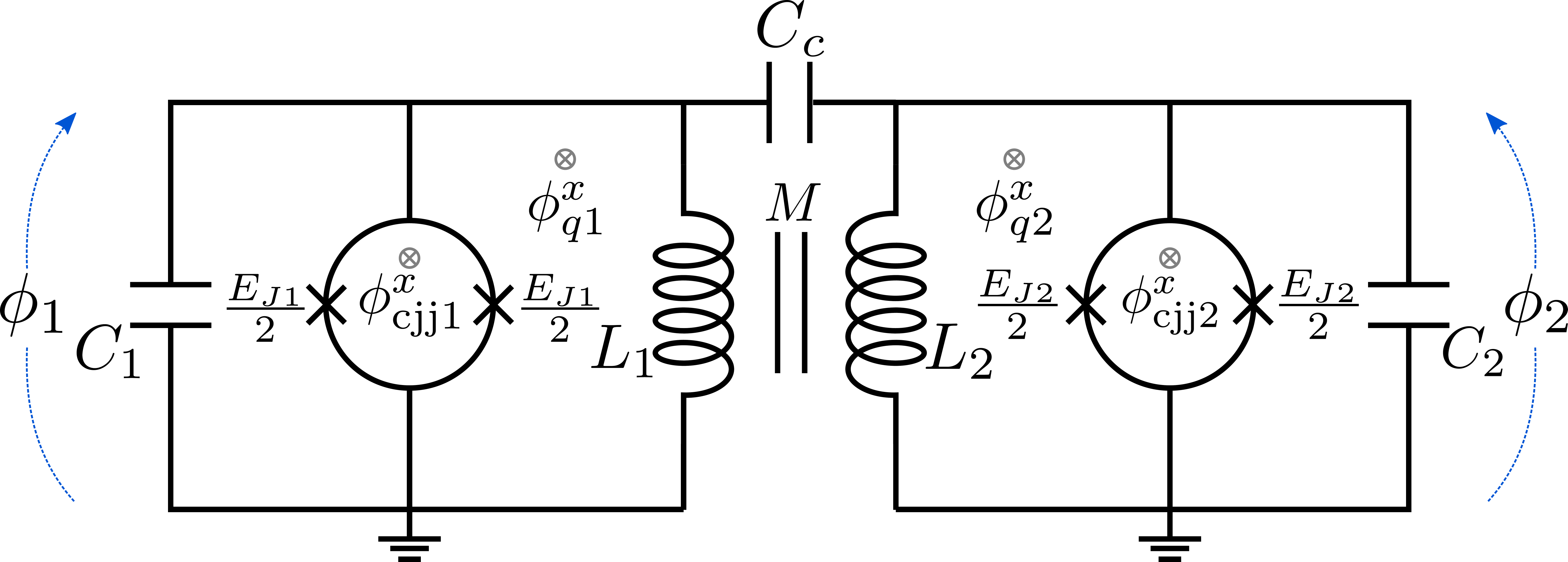}
\caption{Electric circuit of two capacitively and inductively coupled flux qubits.}
\label{fig::fqcoup}
\end{figure}
We consider a system of coupled flux qubits shown in Fig. \ref{fig::fqcoup}. We show why the reduced Hamiltonian obtained using the standard flux qubit basis is, in a certain parameter regime, non-stoquastic even if we allow for single-qubit unitary transformations.

The Hamiltonian of the circuit in Fig. \ref{fig::fqcoup} can be written as 
\begin{multline}\label{eq::hc_mas}
H = \sum_{k=1}^2 4 E_{C k} \hat{q}_k^2 + \frac{1}{2} E_{Lk} \hat{\phi}_k^2 - E_{Jk}^{\mathrm{eff}}  \cos \bigl (\hat{\phi}_k + \phi_{qk}^x \bigr) \\+ 8 E_{C12} \hat{q}_1 \hat{q}_2 + E_{L12} \hat{\phi}_1 \hat{\phi}_2,
\end{multline}
where $E_{Ck}$ and $E_{C12}$ are the diagonal and off-diagonal entries of the charging matrix, respectively. The charging energy matrix is in turn directly related via Eq. \eqref{eq:charg_en} to the capacitance matrix
\begin{equation}
\bm{C} = \begin{bmatrix}
C_1 + C_c & - C_c \\
-C_c & C_2 + C_c
\end{bmatrix}.
\end{equation}
In addition, the $E_{Lk}$ and $E_{L12}$ are the diagonal and off-diagonal entries of the inductive energy matrix, respectively, while $E_{Jk}^{\mathrm{eff}}\bigl(\phi_{\mathrm{cjj}}^x \bigr)$ is the effective Josephson energy defined in Eq.~\eqref{eq::jeff}. We assume that $E_{C12}$ and $E_{L12}$ are much smaller than the gap of the qubit subspace and any other energy level, so that we can treat the coupling at lowest-order perturbation theory. For concreteness, considering a flux qubit with parameters as in Table \ref{tab:params} in the symmetric configuration this gap is $5.9 \, \mathrm{GHz}$.

We can immediately note that if we project the Hamiltonian in Eq.~\eqref{eq::hc_mas} onto a tensor product of qubit spaces, each qubit space associated with an uncoupled Hamiltonian with conjugate variables $q_k, \phi_k$, the presence of both inductive and capacitive couplings will typically lead to Pauli interactions of rank equal to 2. Said differently, the projected two-qubit interaction term of the form $H=\sum_{i=1,j=1}^{3,3} \beta_{ij} P_i \otimes P_j$ is such that the $3 \times 3$ matrix $\bm{\beta}$ has rank 2. 

But when either one of the couplings $E_{C12}$ or $E_{L12}$ is zero, we note that the rank of matrix $\bm{\beta}$ is 1. It can be proved quite directly that a two-qubit Hamiltonian with a rank 1 $\bm{\beta}$-matrix and arbitrary single-qubit terms can be locally sign-cured \cite{klassenTerhal}.

Let us look at this in detail here. We can project onto the qubit subspace for each flux qubit as in Appendix \ref{app::fq} we obtain
\begin{multline}\label{eq::h2qns}
H_{2q}/h = - \frac{\Delta_1}{2} X_1 -  \frac{\varepsilon_1}{2} Z_1 - \frac{\Delta_2}{2} X_2 -  \frac{\varepsilon_2}{2} Z_2 \\
+ J_{YY} Y_1 Y_2 + J_{ZZ} Z_1 Z_2, 
\end{multline}
where
\begin{subequations}\label{eq::jyy}
\begin{align}
J_{YY} &= 8 \frac{E_{C12}}{h} \braket{0| \hat{q}_1|1}_{1} \braket{0| \hat{q}_2|1}_{2}, \\
J_{ZZ} &= \frac{E_{L12}}{h} \braket{0| \hat{\phi}_1|0}_{1} \braket{0| \hat{\phi}_2|0}_{2}.
\end{align}
\end{subequations}
By definition, the tunnel couplings $\Delta_{1,2}$ are positive, see Eq.~\eqref{eq::tcoup}, while the local fields $\varepsilon_{1,2}$, see Eq.~\eqref{eq::eps}, are real. 

If we do not have capacitive coupling $J_{YY}=0$, the Hamiltonian in Eq. ~\eqref{eq::h2qns} is that of a TIM, which can always be made stoquastic via single-qubit unitaries. If the inductive coupling is absent, then $J_{ZZ}=0$. The Hamiltonian is clearly non-stoquastic in the chosen basis as the $Y \otimes Y$ matrix has alternating signs on the off-diagonal elements. However, as said above, we can always make it stoquastic via single-qubit unitaries in the following way. In this case, we first perform a rotation around the $Y$-axis, leaving the term $YY$ unchanged, on each qubit that transforms $- \Delta_k X_{k}/2 -  \varepsilon_{k} Z_k/2 \mapsto - \tilde{\Delta}_{k} X_k/2$, $k=1,2$.
Now we can easily make the Hamiltonian stoquastic by performing the transformation $Y_{1, 2} \leftrightarrow Z_{1, 2}$.

More generally, we show in Appendix \ref{app::paritysym} that if the local fields $\varepsilon_{1,2}$ are zero, even in the case in which we have both capacitive and inductive couplings, the two-qubit Hamiltonian can always be made stoquastic and this in fact holds at arbitrary order in SW perturbation theory.

If the local fields $\varepsilon_{1,2}$ are non-zero, and we have capacitive and inductive coupling such that $\lvert J_{YY} \rvert > \lvert J_{ZZ} \rvert >0  $ we conclude that the Hamiltonian cannot be made stoquastic by a product of two single-qubit unitaries, following the reasoning in Ref.~\cite{klassenTerhal},  The basic (rough) idea is that in order for $H$ to have non-positive off-diagonal elements a term like $Y_1 Y_2$ should be accompanied by a term $X_1 X_2$ of equal magnitude (which it is not) or be rotated away to the $XZ$-plane. In the latter case, one however also rotates the single-qubit $X$ and $Z$ terms into having a $Y$ component, making the Hamiltonian complex and non-stoquastic. 

We remark that, for simplicity of exposition, we present the discussion assuming the validity of the projection, in order to highlight the mechanism that leads to a non-stoquastic behaviour. By refining the perturbation theory, \i.e., using higher order SW transformation for instance \cite{Consani_2020}, the Hamiltonian can still be non-stoquastic {\em even in the absence of an inductive coupling}. In particular, it is shown in Ref. \cite{ozfidan2020} that in the case of strong capacitive coupling, the higher levels of the flux qubits generate an additional $X_1 X_2$ term which can make the Hamiltonian non-stoquastic under single-qubit unitaries.

Naturally, for a two-qubit Hamiltonian we can apply a two-qubit unitary basis change to diagonalize the Hamiltonian, hence there is always a basis change which removes the sign problem. However, one can readily extend this two-qubit case to a line of $N$ coupled flux qubits. When the capacitive and inductive couplings are sufficiently weak (so that $\bm{E_C}$ couples only nearest-neighbor qubits on the line), we obtain a $N$ flux qubit Hamiltonian with $Y_i Y_{i+1}, Z_i Z_{i+1}$ coupling between nearest-neighbor qubits $i$ and $i+1$ on the line. The same arguments then apply as in the two-qubit case: when $\Delta_i \neq 0$, $\epsilon_i \neq 0$ the $N$-qubit Hamiltonian cannot be made stoquastic by a product of $N$ single-qubit basis changes. 

This is thus in sharp contrast with the fact that the general master Hamiltonian in Eq.~\eqref{eq::h_gen} can be called stoquastic in the flux basis and was amenable to the PIMC method, see the numerics in Sec. \ref{subsec:mcs}.

Note also that when coupled flux qubits are non-identical in their parameters, finding a curing transformation for a single pair of qubits does not necessarily imply the existence of a curing transformation which works for the entire set of qubits as the local basis changes have to be chosen to work for each two-qubit interaction.

In the previous discussion, we have shown that the effective qubit Hamiltonian of two coupled flux qubits can be non-stoquastic even after single-qubit unitary rotations. However, there could be other ways to cure non-stoquasticity. 
We show in the next Sec. \ref{app::avoidStoq} that if the capacitive and inductive couplings are weak enough, effective flux qubit Hamiltonians can always be made approximately stoquastic if we perform a canonical transformation before obtaining the reduced qubit Hamiltonian. While these transformations are highly non-local, they can still be implemented efficiently before reducing to a qubit model. In particular, we show that the addition of capacitive couplings to flux qubit Hamiltonians yields, to lowest perturbative approximation, a TIM with modified parameters.  While this derivation relies on the fact that the coupling is weak, it has the appealing feature that it is valid for an arbitrary number of qubits.

\subsection{Flux qubits with weak-strength capacitive and inductive couplings}\label{app::avoidStoq}

We consider flux-qubit systems where the Hamiltonian takes the following particular form
\begin{multline}\label{eq::hcfqi}
H = 4 \hat{\bm{q}}^T \bm{E_C} \hat{\bm{q}} + \frac{1}{2} \hat{\bm{\phi}}^T \bm{E_L} \hat{\bm{\phi}}\\
 -\sum_{k=1}^N E_{Jk}^{\mathrm{eff}} \cos \bigl( \hat{\phi}_k + \phi_{q k}^{x} \bigr) ,
\end{multline}
with  $E_{Jk}^{\mathrm{eff}}$ the effective Josephson energy and $\phi_{q k}^{x}$ the external flux threading the loop formed by the SQUID loop and the corresponding shunting inductance of the $k$th flux qubit.  The inductive energy matrix equals
\begin{equation}
\bm{E_L} = \frac{\Phi_0^2}{4 \pi^2} \bm{L}^{-1},
\end{equation}
with $\bm{L}^{-1}$ the inverse of the inductance matrix ${\bm L}$ (which can be assumed to be positive-definite). 

By taking parameters such that all degrees of freedom are in the flux qubit regime, the Hamiltonian in Eq.~\eqref{eq::hcfqi} models a system of capacitively and inductively coupled flux qubits, where the inductive coupling is expressed in $\bm{E_L}$ and the capacitive coupling is expressed in $\bm{E_C}$. See Fig.~\ref{fig::fqcoup} for two such coupled qubits. \par

We now show how this Hamiltonian, which has no sign problem in the flux basis as we discussed in Sec. \ref{sec:t-reversal}, can be reduced to an effective qubit Hamiltonian which is also stoquastic if the capacitive couplings are small, \i.e., the off-diagonal elements of $\bm{E_C}$ are much smaller than the diagonal ones. In addition, the mutual-inductive couplings between the flux qubits should also be small so that a projection onto the eigenbasis of the uncoupled qubits is a good approximation. 

In some sense this is not a surprising result as our symplectic transformation removes the capacitive couplings, leaving only the inductive couplings which lead, when projected, to rank-1 $\beta$-matrices. \par

We introduce the following canonical transformation 
\begin{equation}
\hat{\bm{q}}' = \bm{S} \hat{\bm{q}}, \quad \hat{\bm{\phi}}' = \bm{S}^{-1} \hat{\bm{\phi}},
\end{equation}
where we defined the matrix
\begin{equation}
\bm{S} = \bm{S}^T=\biggl(\frac{\bm{E_C}}{E_{C0}} \biggr)^{1/2},
\end{equation}
with $E_{C0}$ an arbitrary charging energy which just ensures the entries in $\bm{S}$ are dimensionless. $\bm{S}$ preserves the canonical commutation relations as $\bm{S}=\bm{S}^T$, i,e. $[\hat{\phi}'_k,\hat{q}_l']=[\hat{\phi}_k,\hat{q}_k]=i \delta_{kl}$.
We will drop the primes from now on for these canonical variables.
The Hamiltonian in Eq.~\eqref{eq::hcfqi} becomes
\begin{multline}\label{eq::hfqci2}
H = 4 E_{C0} \hat{\bm{q}}^T \hat{\bm{q}} +  \frac{1}{2} \hat{\bm{\phi}}^T \bm{E_L}' \hat{\bm{\phi}} - \\
\sum_{k=1}^N E_{Jk}^{\mathrm{eff}} \cos \biggl[S_{kk} \hat{\phi}_k 
+ \phi_{q k}^{x} + \underset{l \neq k}{\sum_{l = 1}^N} S_{k l} \hat{\phi}_{l}  \biggr],
\end{multline}
where we have introduced an effective inductive energy matrix 
\begin{equation}
\bm{E_L}' = \bm{S}^T \bm{E_L} \bm{S}.
\end{equation}

We now first show that we can map the Hamiltonian in Eq.~ \eqref{eq::hfqci2} to a transverse field Ising model when the capacitive couplings between the flux qubits are not too large. This implies that matrix $\bm{E_C} \propto \bm{C}^{-1}$ has off-diagonal elements which are small compared to its diagonal elements and hence so will $\bm{S}$ when we treat the capacitive coupling between flux qubits as a perturbation.

In this case, we can expand each cosine term as
\begin{multline}
-E_{Jk}^{\mathrm{eff}} \cos \biggl[S_{k k} \hat{\phi}_k 
+ \phi_{q k}^{x} + \underset{l \neq k}{\sum_{l = 1}^N}  S_{k l} \hat{\phi}_{l}  \biggr] \\ \approx 
-E_{Jk}^{\mathrm{eff}} \cos \bigl[S_{k k} \hat{\phi}_k 
+ \phi_{q k}^{x} \bigr] + \\
E_{Jk}^{\mathrm{eff}} \sin \bigl[S_{k k} \hat{\phi}_k 
+ \phi_{q k}^{x} \bigr] \underset{l \neq k}{\sum_{l = 1}^N}  S_{k l} \hat{\phi}_{l} .
\end{multline}
This allows us to rewrite Eq.~\eqref{eq::hfqci2} as 
\begin{equation}
H \approx \sum_{k=1}^N H_k + H_{c}^{\mathrm{ind}} + H_{c}^{\mathrm{jj}},
\end{equation}
where we defined 
\begin{itemize}
\item the effective Hamiltonian of the $k$th flux qubit $H_k$ as 
\begin{multline}\label{eq::hk_eff}
H_k = 4 E_{C0} \hat{q}_k^2 + \frac{(\bm{E_L}')_{kk}}{2} \hat{\phi}_k^2 \\
- E_{Jk}^{\mathrm{eff}} \cos \bigl[S_{k k} \hat{\phi}_k
+ \phi_{q k}^{x} \bigr];
\end{multline}
\item the inductive coupling Hamiltonian $H_{c}^{\mathrm{ind}}$ as 
\begin{equation}
H_{c}^{\mathrm{ind}} = \sum_{\langle k, l \rangle} (\bm{E_{L}}')_{kl} \hat{\phi}_k \hat{\phi}_l;
\end{equation}
\item the additional coupling due to the Josephson junctions $H_{c}^{\mathrm{jj}}$ as 
\begin{equation}\label{eq::hcjj}
H_{c}^{\mathrm{jj}} = \sum_{k=1}^N E_{Jk}^{\mathrm{eff}}  \sin \bigl[S_{k k} \hat{\phi}_k 
+ \phi_{q k}^{x} \bigr]  \underset{l \neq k}{\sum_{l = 1}^N}  S_{k l} \hat{\phi}_{l}.
\end{equation}
\end{itemize}

We can now use the flux qubit Hamiltonians $H_k$ to define a local computational basis $\mathcal{B}_k =\{\ket{0}_k, \ket{1}_k \}$, similar to what is done in Appendix \ref{app::fq}. The global computational basis $\mathcal{B}$ is obtained by taking all possible tensor products of these states, \i.e., $\mathcal{B} = \bigotimes_{k=1}^N \mathcal{B}_k$. By projecting onto this basis we obtain a reduced $N$-qubit Hamiltonian that can be written as
\begin{equation}\label{eq::htimeff}
H_{\mathrm{eff}}/h = -\biggl( \sum_{k=1}^N \frac{\Delta_k}{2} X_k + \frac{\varepsilon_k}{2} Z_k \biggr) + 
\sum_{\langle k, l \rangle} J_{kl} Z_k Z_l.
\end{equation}
The parameters in this Hamiltonian are obtained as follows. With the definition of the flux qubit Hamiltonian $H_k$ defined in Eq. \eqref{eq::hk_eff} the tunnel couplings $\Delta_k$ are given by
\begin{equation}
\Delta_k = \frac{E_e^{(k)}-E_{g}^{(k)}}{h},
\end{equation}
with $E_{g, e}^{(k)}$ are the ground and first-excited eigenenergies of $H_k$ in the double well configuration $\phi_{qk}^x=\pi$. By defining $\delta_{qk}^x = \phi_{qk}^x - \pi$ the parameters $\varepsilon_k$ and considering small $\delta_{qk}^x$, similarly to Appendix \ref{app::fq}, we define
\begin{multline}
\varepsilon_k = \frac{E_J^{\mathrm{eff}}}{h} \delta_{qk}^x \bigl[\braket{0 | \sin \bigl(S_{kk} \hat{ \phi}_k \bigr) | 0}_k -  \braket{1 | \sin \hat{\phi}_k| 1}_k \bigr] = \\
2 \frac{E_J^{\mathrm{eff}}}{h} \delta_{qk}^x \braket{0 | \sin \bigl(S_{kk} \hat{\phi}_k \bigr)| 0}_k.
\end{multline}
Finally, neglecting the small corrections we get when $\delta_{qk}^x \neq 0$, the exchange coupling $J_{kl}$ reads
\begin{multline}
J_{kl} = \frac{E_{Lkl}'}{h}\braket{0 | \hat{\phi}_k | 0}_k \braket{0 | \hat{\phi}_l | 0}_l - \\ \frac{E_J^{\mathrm{eff}}}{h} S_{kl} \braket{0 | \sin\bigl(S_{kk}\hat{\phi}_k\bigr) | 0}_k    \braket{0 | \hat{\phi}_l | 0}_l.
\end{multline}

Eq.~\eqref{eq::htimeff} is the Hamiltonian of a TIM, which is stoquastic in the computational basis. Notice that this derivation is valid for an arbitrary number of flux qubits. 

The transverse field Ising model in Eq.~\eqref{eq::htimeff} can also be mapped to a classical system and be studied using the PIMC method. We refer the reader to Ref. \cite{martonakSantoro} for a derivation. The $N+1$-dimensional classical Hamiltonian associated with the TIM reads
\begin{multline}
\mathcal{H}_{\mathrm{eff}, c}/h = -\sum_{s=1}^M \biggl( \sum_{k=1}^N J_k^{\perp} \sigma_{k}^{(s)} \sigma_{k}^{(s+1)} + \sum_{k=1}^N \frac{\varepsilon_k}{2} \sigma_{k}^{(s)} - \\
\sum_{\langle k, l \rangle} J_{kl} \sigma_k^{(s)} \sigma_l^{(s)} \biggr),
\end{multline}
where the variables $\sigma_{k}^{(s)}$ are classical spins which can take value $\pm 1$ and we defined the parameter
\begin{equation}
J_k^{\perp} = -\frac{M}{2 \beta} \ln \tanh \frac{\Delta_k \beta}{2 M},
\end{equation}
with $M$ the number of Trotter slices as in Sec. \ref{sec::cqedflux}.

\subsection{Monte Carlo simulations}\label{subsec:mcs}
In this subsection we perform sign-problem-free Monte Carlo simulations for the average thermal energy of a system of two capacitively coupled flux qubits. We begin by studying the problem using the path integral representation of the original Hamiltonian in the flux basis as discussed in Sec. \ref{sec:t-reversal} for the case of weak capacitive coupling. We compare the result with the PIMC with those using the effective TIM discussed in Sec. \ref{app::avoidStoq}.  The goal is to compare the results using these two methods for weak capacitive coupling versus the exact results for the estimation of the average thermal energy
\begin{equation}
\langle H \rangle_{\beta} = \frac{1}{Z}\mathrm{Tr} \biggl(H e^{- \beta H} \biggr),
\end{equation}
with $Z = \mathrm{Tr}[ \exp(-\beta H) ]$ the partition function. We also study whether in the case of strong capacitive coupling, for which the effective qubit Hamiltonian is non-stoquastic as in Ref.~\cite{ozfidan2020}, the PIMC using the original Hamiltonian provides reliable results.
We consider two identical flux qubits with parameters as in Table \ref{tab:params}. Throughout this subsection the minimum of the potential is taken as the zero of the energy in any parameter set.

\par 
\begin{figure}
    \centering
    \includegraphics[scale=0.4]{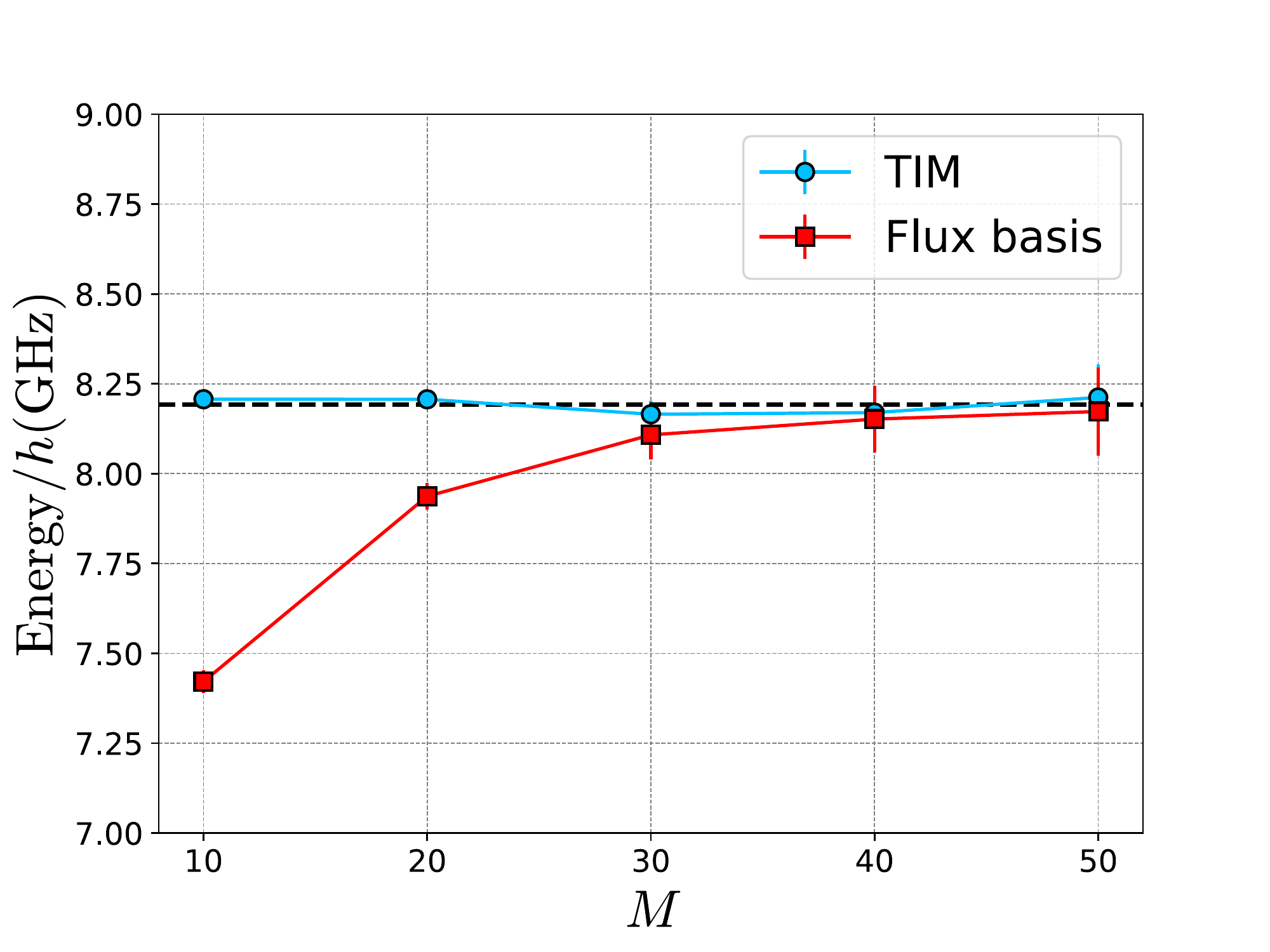}
    \caption{Average thermal energy using the PIMC in the flux basis and the effective TIM model obtained in Sec. \ref{app::avoidStoq} as a function of the number of Trotter steps $M$. The temperature is taken to be $(h \beta)^{-1} = 0.93 \, \mathrm{GHz}$. The coupling capacitance is chosen to be $C_c = 10 \, \mathrm{fF}$, so that we obtain $E_{C12}/h = 0.008 \, \mathrm{GHz} \ll E_C/h$. Both flux qubits are operated in the symmetric double well configuration with $\phi_{q 1, 2}^x = \pi$. The black dashed line corresponds to the exact thermal energy obtained from numerical diagonalization, while the ground-state energy is $E_{g} = 7.675 \, \mathrm{GHz}$.}
    \label{fig:qmc_res}
\end{figure}

\begin{figure}
\centering
    \includegraphics[scale=0.4]{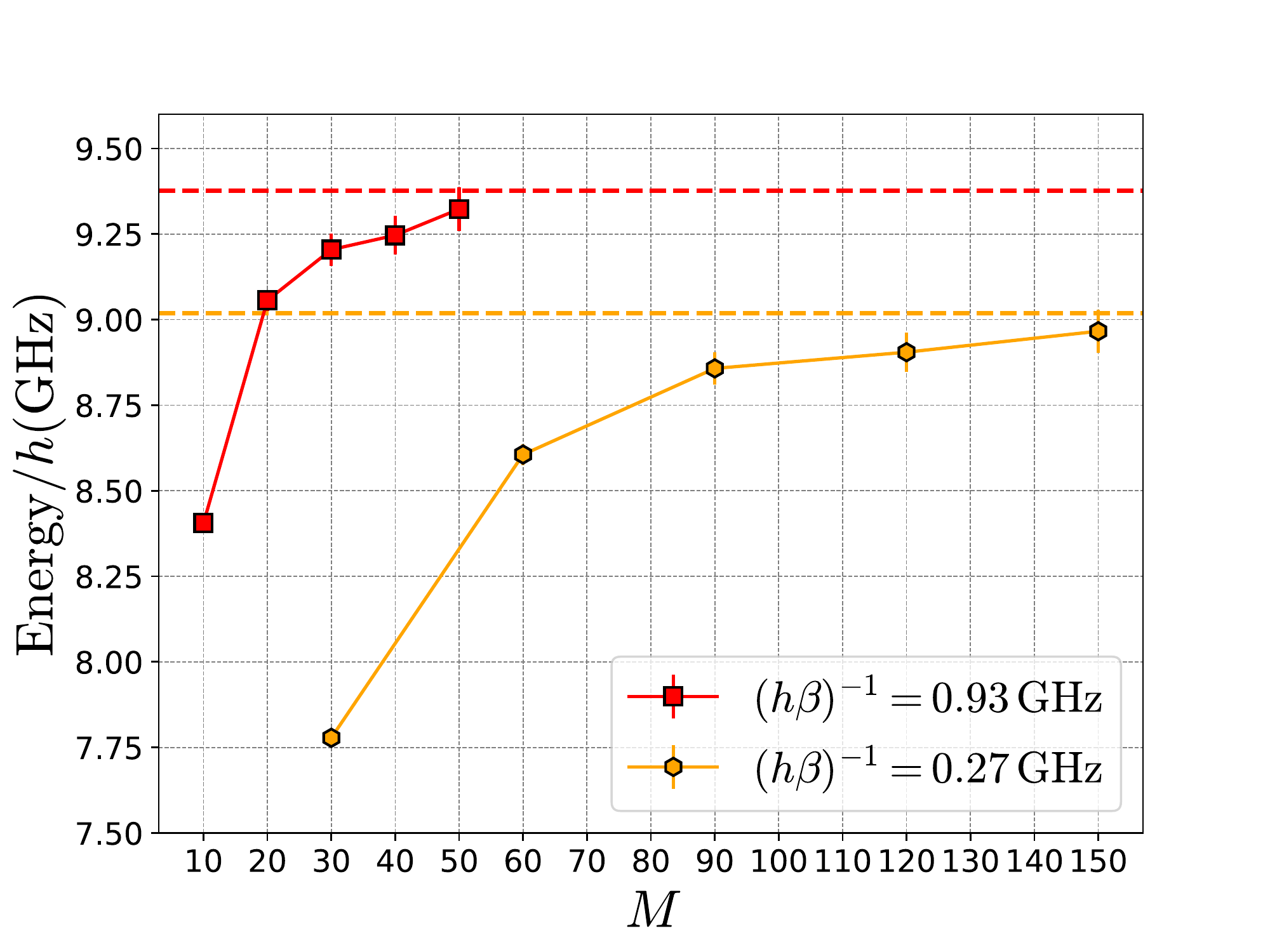}
    \caption{Average thermal energy using the PIMC in the flux basis for the case of strong capacitive coupling. The coupling capacitance is $C_c= 104 \, \mathrm{fF}$ giving $E_{C12}/h = 0.062 \, \mathrm{GHz}$. The flux qubits are operated in an asymmetric configuration with $\delta_{q1}^x/2 \pi = 10^{-4}$, $\delta_{q2}^x/2 \pi = 2\times10^{-4}$. According to the color coding, the dashed lines denote the exact thermal energy at the respective temperature. The ground-state energy is $E_g/h = 9.016 \, \mathrm{GHz}$, which is within $0.2 \%$ accuracy the thermal energy in the orange (light gray) dashed line.}
    \label{fig:qmc_cc1_tilt}
\end{figure}

\begin{table}[h]
   \centering
   \begin{tabular}{c|c}
   \textbf{Parameter} & $\mathrm{GHz}$ \\ 
   \hline
	$E_C/h $ 						& $0.124 $ 	\\ 
	$E_J/h$						& $1600$		\\
	$E_J^{\mathrm{eff}}/h$		& $760$		\\
	$E_L/h$						& $704$  	
   \end{tabular}
   \caption{Parameters for simulations. The effective Josephson energy is obtained by setting the external fluxes to $\phi_{\mathrm{cjj}}^x = 0.685550 \times \pi$. This choice of parameters corresponds to $E_J^{\mathrm{eff}}/E_{L}=1.08$ and the tunnel coupling $\Delta/h = (E_e - E_g)/h = 1.36 \, \mathrm{GHz}$}
   \label{tab:params}
\end{table}

We start by considering the case of a symmetric potential for both flux qubits and small coupling capacitance. We provide some details of the Monte Carlo simulations in Appendix \ref{app::detmc}. The results are shown in Fig. \ref{fig:qmc_res}. We see that the PIMC in the flux basis needs more Trotter slices to accurately estimate the average thermal energy compared to the TIM. Qualitatively this happens because in the TIM we are using a basis in which the Hamiltonian is approximately diagonal and so the quantum effects are already taken into account. Notice also that the simulations in the flux basis consistently underestimate the average thermal energy for small $M$. This is due to the fact that the Trotter break-up formula neglects the commutation relation, \i.e., quantum mechanical effects, and so we expect to have lower zero-point energy compared to the exact quantum solution for small $M$. However, we see that as expected with $50$ Trotter slices, PIMC in the flux basis accurately estimates the average thermal energy. \par 
The PIMC method in the flux basis can also be used to study the case of strong capacitive coupling as considered in Ref. \cite{ozfidan2020}, without fundamental limitations. This is shown in Fig. \ref{fig:qmc_cc1_tilt}. For the given parameters, the effective qubit Hamiltonian is non-stoquastic as discussed in Subsec. \ref{sec::example}. However, the PIMC using the original Hamiltonian is still able to estimate the average thermal energy accurately with similar number of Trotter slices for the same temperature as for the case of Fig. \ref{fig:qmc_res}. However, for this case the Trotter error is clearly worse as we can see from the fact that at low $M$ we have larger relative error compared to the case of weak coupling. This is simply due to the larger coupling capacitance and not a signature of a fundamental obstruction. In addition, the orange (light gray) line also shows that similar accuracy can be achieved even if we reduce the temperature so that the thermal energy gets closer to the ground-state energy. This naturally comes at the price of increasing the number of Trotter slices by a factor of three, while the number of Metropolis iterations was not changed between the two different temperatures.

\section{Discussion}\label{sec::conclusions}

In this paper we have seen that qubit Hamiltonians which may appear to be non-stoquastic can have `master' circuit QED Hamiltonians which are manifestly stoquastic. We have used this observation to propose an efficient simulation method for quantum adiabatic computation with such Hamiltonians, using path integral Monte Carlo methods.

It is not entirely straightforward to reconcile the projected non-stoquastic Hamiltonian in Sec. \ref{sec::example} with the TIM Hamiltonian in \ref{sec::stoqEff} as we do not discuss the error induced by only using the lowest-order Schrieffer-Wolff projection. For example, if the non-stoquasticity of Eq.~\eqref{eq::h2qns} is of the same order of magnitude as the error induced by perturbation theory, then one cannot draw any hard conclusions.

We have observed that circuit QED Hamiltonians are generically stoquastic and thus amenable to Monte Carlo methods in their continuous variable representation if they don't contain explicit time-reversal invariance breaking terms due to driving or non-reciprocity in the electric circuit. We have also recently become aware of Ref. \cite{halversonHen2020}, where similar conclusions are drawn, although using a different Monte Carlo method that requires the discretization of the flux degrees of freedom. In our case, instead, we do not require this flux discretization, but we rely on a finite number of Trotter slices. 

Naturally, these arguments do not directly apply to fermionic systems or fermionic field theories, in which the path integral is an integral over (non-commuting) Grassmann variables. Alternatively, for fermions treated in first quantization, we can view the sign problem as arising from the fact that we are restricting the space of states to wave-functions which are fully anti-symmetric under the interchange of particles: a Gibbs state or a ground state in the full phase space is not the relevant physical object to study. If we use second quantization, we encapsulate the anti-symmetry constraint, --working in the subspace of anti-symmetric wavefunctions--, and generally see that the corresponding Hamiltonian, say a Hubbard model, is not stoquastic when expressed in a fermionic Fock or qubit basis. Time-reversal does however play a role in some special cases when we avoid the sign problem for fermionic systems \cite{Wu_2005}.

As for complexity, it is important to note that it is highly unlikely that one can find computationally-efficient curing transformations which map any (local) Hamiltonian onto a stoquastic Hamiltonian as it would have unlikely complexity-theoretic consequences. It was shown in Ref.~\cite{bravyiTerhalDV} that the ground-state energy estimation problem for stoquastic qubit Hamiltonians is a problem contained in ${\rm AM}$ (and ${\rm StoqMA} \subseteq {\rm AM}$). The class AM is contained in the so-called polynomial hierarchy, while on the other hand, BQP, let alone QMA, is not believed to be contained in the polynomial hierarchy \cite{RT:BQP,aaronson:BQP}.
Establishing the precise physical origin of the sign problem, and when it can be avoided, is important as the sign problem is precisely what separates quantum from classical computation: At least, the sign problem necessitates the use of quasi-probability distributions (which lead to potentially-exponential variances in Monte Carlo simulations) which can be used quite widely for the simulation of quantum computation by classical stochastic means \cite{pashayan2015, howardCampbell,  mariEisert}. 

\section{Acknowledgements}

We thank David DiVincenzo and Joel Klassen for many insightful discussions on the topic of this paper. We thank Marios Ioannou and Sergey Bravyi for discussions on perturbative gadgets. We thank Gioele Consani for useful discussions about the SW transformation for coupled flux qubits. We thank Adrian Parra-Rodriguez for a useful comment on canonical transformations in the presence of non-reciprocal elements. We are also thankful to Jan Reiner for an important remark on single-qubit unitaries curing non-stoquasticity.  Our work was supported by ERC grant EQEC No. 682726.

\appendix

\section{PIMC in the flux basis}\label{app::pimc_flux}
\numberwithin{equation}{section}

In this Appendix we explicitly derive the PIMC method for the general Hamiltonian in Eq.~\eqref{eq::h_gen1} by performing a mapping to a classical model and its partition function. The derivation is a simple adaptation of those that can be found in Refs. \cite{landauBinder, ceperley1995}, where the only additional complication that is added is that the inverse of the capacitance matrix in Eq.~\eqref{eq::h_gen} is not diagonal. We consider the general task of computing the thermal average of an observable $O$:
\begin{equation}\label{eq::o_avg}
\langle O \rangle_{\beta} = \frac{1}{Z} \mathrm{Tr} \biggl ( O e^{-\beta H} \biggr).
\end{equation}
We will evaluate the trace in the flux $\ket{\bm{\Phi}}$ basis and we will further assume that the observable $O$ that we are evaluating is diagonal in this basis. We remark that one can also evaluate the thermal average of $H$ itself, even though it has an off-diagonal kinetic term in the flux basis. This follows by virtue of the quantum virial theorem \cite{fockVirial}, which states that that the average of the kinetic energy $K$ in any eigenstate of $H$, and thus also for thermal averages, satisfies
\begin{equation}
\langle K \rangle_{\beta} = \frac{1}{2} \sum_{k=1}^N \biggl \langle \hat{\Phi}_k \frac{\partial U}{\partial \Phi_k} \biggr \rangle_{\beta}.
\end{equation}
We make use of this result to evaluate the thermal energies for the PIMC in the flux basis discussed in Subsec. \ref{subsec:mcs}. 

For general off-diagonal observables $O$ there is no rigorous relation between $\langle O \rangle_{\beta}$ and the evaluation of $Z$.

Let us start by rewriting Eq. \eqref{eq::o_avg} as
\begin{equation}
\langle O \rangle_{\beta} \approx \frac{1}{Z} \int d \bm{\Phi}_1 \braket{\bm{\Phi}_1| O \underbrace{e^{-\beta H/M} \dots e^{-\beta H/M}}_{M \, \text{times}} | \bm{\Phi}_1},
\end{equation}
where the integral is over $\mathbb{R}^N$ and we compactly denote $d \bm{\Phi}_1 = \prod_{k=1}^N d \Phi_{1k}$. 
Assuming $\beta/M \ll 1$ we can use Trotter's break-up formula \cite{suzuki1976} and approximate
\begin{equation}
e^{-\beta H/M} = e^{-\beta (K + U)/M} \approx e^{-\beta K/M} e^{-\beta U/M}.
\end{equation}
Inserting the identity
\begin{equation}\label{eq::id}
\int d \bm{\phi} \ket{\bm{\Phi}}\bra{\bm{\Phi}} = \int d \bm{Q} \ket{\bm{Q}}\bra{\bm{Q}} = \mathds{1},
\end{equation}
in the flux basis $M$ times we obtain
\begin{multline} \label{eq::o_avg2}
\langle O \rangle_{\beta} \approx \frac{1}{Z} \int d \bm{\Phi}_1  d \bm{\Phi}_2 \dots d \bm{\Phi}_{M} \braket{\bm{\Phi}_1| O | \bm{\Phi}_1} \times \\ \braket{\bm{\Phi}_1| e^{-\beta K/M} | \bm{\Phi}_M}\dots  \braket{\bm{\Phi}_{2}| e^{-\beta K/M}| \bm{\Phi}_1}\times \\
e^{-\beta/M \sum_{s=1}^{M}U(\bm{\Phi}_s)},  
\end{multline}
where we used the fact that $O$ is diagonal in the flux basis. We thus need to evaluate the matrix element $\bra{\bm{\Phi}_{s+1}} e^{-\beta K/M} \ket{\bm{\Phi}_s}$. Using Eq. \eqref{eq::id} in the charge basis and \cite{sakurai}
\begin{equation}
\braket{\bm{\Phi} |\bm{Q}} = \frac{1}{(2 \pi \hbar)^{N/2}} e^{i \bm{Q}^T \bm{\Phi}/\hbar},
\end{equation}
we obtain
\begin{multline}\label{eq::four}
\braket{\bm{\Phi}_{s+1} |e^{-\beta K/M} | \bm{\Phi}_{s}} = \bra{\bm{\Phi}_s} e^{-\frac{\beta}{2 M} \bm{\hat{Q}}^T  \bm{C}^{-1} \bm{\hat{Q}}} \ket{\bm{\Phi}_{s+1}} = \\
.\frac{1}{(2 \pi \hbar)^N} \int d \bm{Q} e^{\frac{i}{\hbar} \bm{Q}^{T}(\bm{\Phi}_{s+1}- \bm{\Phi}_{s})} e^{- \frac{\beta}{2 M} \bm{Q}^{T} \bm{C}^{-1} \bm{Q}} = \\
\sqrt{\det \bm{C}} \biggl( \frac{M}{\hbar^2 2 \pi \beta} \biggr)^{\frac{N}{2}} \exp \biggl \{- \frac{M}{2 \hbar^2 \beta} \lvert \bm{C}^{1/2}(\bm{\Phi}_{s+1} - \bm{\Phi}_{s}) \rvert^2 \biggr \},
\end{multline}
which is clearly positive. Eq. \eqref{eq::o_avg2} becomes
\begin{equation}\label{eq::o_avg_pi}
\langle O \rangle_{\beta} \approx \int  d \bm{\Phi}_1  d \bm{\Phi}_2 \dots d \bm{\Phi}_{M} \braket{\bm{\Phi}_1| O | \bm{\Phi}_1} p(\bm{\Phi}_1, \dots, \bm{\Phi}_M),
\end{equation}

where we defined the path probabilities
\begin{multline}\label{eq::path_p}
p(\bm{\Phi}_1, \dots, \bm{\Phi}_M) = \frac{w(\bm{\Phi}_1, \dots, \bm{\Phi}_M)}{Z} = \\  \frac{(\det \bm{C})^{\frac{M}{2}}}{Z} \biggl( \frac{M}{\hbar^2 2 \pi \beta} \biggr)^{\frac{N M}{2}} \exp(-\beta \mathcal{H}_c)
\end{multline}
with periodic boundary condition $\bm{\Phi}_{M+1} = \bm{\Phi}_1$, and classical Hamiltonian given in Eq.~\eqref{eq::h_c} in the main text.

Notice that all path probabilities are positive and they are correctly normalized since by repeating the previous derivation we can write the partition function as
\begin{equation}
Z = \int  d \bm{\Phi}_1  d \bm{\Phi}_2 \dots d \bm{\Phi}_{M} w(\bm{\Phi}_1, \dots, \bm{\Phi}_M).
\end{equation}
We thus have written the thermal average of a diagonal operator as the average of an estimator $\braket{\bm{\Phi}_1| O | \bm{\Phi}_1}$ over a classical probability distribution. We remark that also $\sum_{m=1}^M \braket{\bm{\Phi}_m| O | \bm{\Phi}_m}/M$ is a valid, unbiased estimator.  \par 
Eq.~\eqref{eq::o_avg_pi} is the basis for the PIMC method, where we  
sample from the probability distribution over path configurations $p(\bm{\Phi}_1, \dots, \bm{\Phi}_M)$ for instance by using the Metropolis-Hastings algorithm \cite{metropolis, hastingsMetropolis} that we detail in the next Subsection \ref{sec::mm} for completeness.\par

\subsection{Metropolis-Hastings reviewed}
\label{sec::mm}

The Metropolis-Hastings algorithm allows to sample from an arbitrary probability distribution, in our case $p(\bm{\Phi})$, given the ability to compute a function $f(\bm{\Phi})$ proportional to it, \i.e., $f(\bm{\Phi})= c p(\bm{\Phi})$ for some $c \in \mathbb{R}$. 

In our case 
\begin{equation}
\frac{p(\bm{\Phi})}{p(\bm{\Phi}')}=\exp(-\beta(\mathcal{H}_c(\bm{\Phi})-\mathcal{H}_c(\bm{\Phi}')))
\end{equation}

The algorithm works as follows.
\begin{enumerate}
\item Choose an initial configuration $(\bm{\Phi}_1^{(k)}, \dots, \bm{\Phi}_M^{(k)})$, $k=0$. The initial configuration can be chosen randomly, but this is not necessary.
\item Propose a new configuration $(\bm{\Phi}_1', \dots, \bm{\Phi}_M')$ according to some probability distribution (transition rule). Evaluate the variation of the Hamiltonian (energy) $\Delta \mathcal{H}^{k} = \mathcal{H}_c(\bm{\Phi}')-\mathcal{H}_{c}(\bm{\Phi}^{(k)})$. It is assumed that the transition rules are chosen such that the probability for a transition from $(\bm{\Phi}_1, \dots, \bm{\Phi}_M)$ to $(\bm{\Phi}_1', \dots, \bm{\Phi}_M')$ is the same as that of transition from $(\bm{\Phi}_1', \dots, \bm{\Phi}_M')$ to $(\bm{\Phi}_1, \dots, \bm{\Phi}_M)$ (Markov chain is symmetric). 
\item Accept the new configuration and set $(\bm{\Phi}_1^{(k+1)}, \dots, \bm{\Phi}_M^{(k+1)})= (\bm{\Phi}_1', \dots, \bm{\Phi}_M')$ with probability 
\begin{equation}
p = \min \biggl[1, e^{-\beta \Delta \mathcal{H}^k} \biggr],
\end{equation}
otherwise $(\bm{\Phi}_1^{(k+1)}, \dots, \bm{\Phi}_M^{(k+1)})=(\bm{\Phi}_1^{(k)}, \dots, \bm{\Phi}_M^{(k)})$. 
\item Update $k=k+1$ and go to 2.
\item Halt the algorithm when a sufficient number of configurations have been generated from which we can compute the desired averages as arithmetic averages.
\end{enumerate}
We see that the Metropolis-Hastings algorithm generates a Markov chain whose equilibrium distribution can be shown to be the desired probability distribution. Thus, we should start to average only when equilibrium is reached. Also, the performance of the algorithm is strongly influenced by the choice of the transition rule. These can be broadly distinguished into two main categories:
\begin{enumerate}
\item local update: at step $k$ a random particle $s$ with $s=1, \dots, M$ in imaginary time is chosen and its configuration is randomly changed as $\bm{\Phi}_s'= \bm{\Phi}_s^{k} + \bm{\delta}$ where $\bm{\delta}$ is a $N$-dimensional vector of random variables, usually chosen uniformly within a range $[-\Delta, \Delta]$ for some $\Delta$;
\item global update: at step $k$ all particles are shifted by the same $N$-dimensional vector $\bm{\delta}$ of random variables.
\end{enumerate} 
One can also come up with mixed strategies. As pointed out in \cite{ceperley1995} it is generally good to have a variety of update rules that we select with a certain probability. These considerations are however always dependent on the particular system we are dealing with.

\subsection{Details of the Monte Carlo simulations}\label{app::detmc}
We give some details of the Monte Carlo simulations discussed in Sec. \ref{sec::example}. In Fig. \ref{fig:qmc_res}, we initialize in both the PIMC and TIM simulations the corresponding classical system in a random configuration. We let the system equilibrate for $5 \times 10^6$ Metropolis iterations, after which we start to sample the energy every $1000$ iterations. We continue to run the Metropolis algorithm until $30 \times 10^6$ iterations are reached. In the calculation of the error bars we take into account the correction due to the correlation between the samples by explicitly computing the autocorrelation time of the samples. This explains why the error bars are increasing with the number of Trotter slices $M$ in Fig. \ref{fig:qmc_res}, since if we fix the number of iterations, we expect the autocorrelation time to increase with $M$. 
For the PIMC in the flux basis we apply local updates with probability $0.9$, while otherwise we attempt a global update. In both cases, we attempt to modify the chosen flux variables by shifting them by a certain $\delta$ from a uniform distribution in $[-0.75, 0.75]$ (see discussion in the previous subsection). 
The same procedure is applied for Fig. \ref{fig:qmc_cc1_tilt}. A similar update rule is applied for the PIMC derived from the TIM model. With probability $0.9$ we apply a local update where we suggest to flip a random spin. Otherwise, we attempt to flip all spins.

\section{Stoquasticity of the Cooper-pair box}
\label{app:transmon}

The quantum Hamiltonian of a Cooper-pair box or transmon qubit is
\begin{equation}\label{eq::tra}
    H_{\rm transmon}=\frac{1}{2C}(\hat{Q}-Q_g)^2- E_J(\cos(2\pi \hat{\Phi}/\Phi_0),
\end{equation}
with $\Phi_0=\frac{h}{2e}$, as a special case of Eq.~\eqref{eq::shift-H}. The conjugate operators flux $\hat{\Phi}$ and charge $\hat{Q}$ take eigenvalues in $\mathbb{R}$ so that this shifted Hamiltonian can be made manifestly stoquastic in the flux qubit basis by a simple transformation, as discussed in the main text, namely Eq.~\eqref{eq::bs}.

When treating the Hamiltonian in Eq.~\eqref{eq::tra}, one often moves to a rotor basis defined by a $2\pi$-periodic phase $\varphi$ and integer $n\in \mathbb{Z}$ \cite{koch2007}. We can indeed convert from $\hat{\Phi}$ and $\hat{Q}$ to $\hat{\varphi}$ and $\hat{n}$ by defining the basis
\begin{equation}
    \ket{\varphi}=\sum_{k \in \mathbb{Z}} \ket{\phi=\varphi+2\pi k},
\end{equation}
with $\phi=\frac{2\pi \Phi}{\Phi_0}$.
This basis $\ket{\varphi}$ is an eigenbasis for the {\em subspace} of the oscillator space defined by the operator $S_Q=\exp(i \pi\hat{Q}/e)$ taking eigenvalue 1. In this (rotor) subspace we thus have that $\hat{Q}=2 e \hat{n}$ takes eigenvalues $2e n$ with $n \in \mathbb{Z}$, which is interpreted as there being an offset of $n$ Cooper pairs with total charge $2e n$ on the superconducting island defining the transmon qubit.

In this subspace the transmon Hamiltonian of Eq.~\eqref{eq::tra} equals
\begin{equation}\label{eq::transmon}
H_{\rm transmon, sub}=    4 E_C(\hat{n}-n_g)^2-E_J \cos(\hat{\varphi}),
\end{equation}
with offset charge $n_g \in [0,1)$ and $Q_g=2 e n_g$. 
We could have picked another rotor subspace in which $S_Q$ takes the eigenvalue, say, $e^{i 2\pi \tilde{n}_g}$ for some $\tilde{n}_g$. The basis for this subspace is 
\begin{equation}
    \ket{\varphi}_{\tilde{n}_g}=\sum_{k \in \mathbb{Z}} e^{2 \pi i \tilde{n}_g k} \ket{\phi=\varphi+2\pi k}.
\end{equation}
since $S_Q \ket{\varphi}_{\tilde{n}_g}=e^{i 2 \pi \tilde{n}_g}\ket{\varphi}_{\tilde{n}_g}$.

The spectrum and eigenstates of $H_{\rm transmon,sub}$ in Eq.~\eqref{eq::transmon} relate to eigensolutions of the Mathieu equation \cite{koch2007} and depend on $n_g$. For $n_g \neq 0$, the ground-state $\ket{\psi_{0}}=\int_0^{2\pi} d\varphi\; \psi_{0}(\varphi) \ket{\varphi}$ has a complex wavefunction $\psi_0(\varphi)$  \cite{cottet:thesis, koch2007}. For $n_g=0$, the wavefunction $\psi_0(\varphi) \geq 0$. 

We can consider in which subspace the Hamiltonian has a ground-state with minimal energy {\em overall}. We observe that by going to the subspace in which $\tilde{n}_g=n_g$, we obtain a Hamiltonian as in Eq.~\eqref{eq::transmon} with $n_g=0$. The standard spectrum of the transmon qubit \cite{koch2007} shows that this choice achieves the lowest energy eigenvalue. Hence the global ground-state is a non-negative wavefunction in the subspace basis $\ket{\varphi}_{\tilde{n}_g}$. We observe that the basis $\ket{\varphi}_{\tilde{n}_g=n_g}$ is non-negatively related to the transformed basis $\ket{\Phi'}=e^{i \hat{\Phi} Q_g/\hbar} \ket{\Phi}$ in which the original Hamiltonian was explicitly stoquastic: this holds as $e^{i \hat{\Phi}Q_g/\hbar} \ket{\varphi}=\ket{\varphi}_{n_g}$.

Thus we see that the fact that the ground-state wave-function is complex in some rotor subspace is entirely compatible with the stoquasticity of the Hamiltonian (when considered in the full space and in the right basis).

On a separate note, the convergence and accurate predictions of the Monte Carlo path integral simulation of the transmon qubit in the subspace labeled by $\tilde{n}_g$ can be examined. It can depend on whether the numerical simulation varies the winding number or not \cite{HGS:winding}. Here the winding number is the number of times the phase $\varphi$ wraps around $2\pi$ in the path integral.

\section{Flux qubit Hamiltonians}\label{app::hamflux}

\subsection{The flux qubit reviewed}\label{app::fq}

\begin{figure}
\centering
\includegraphics[scale=0.11]{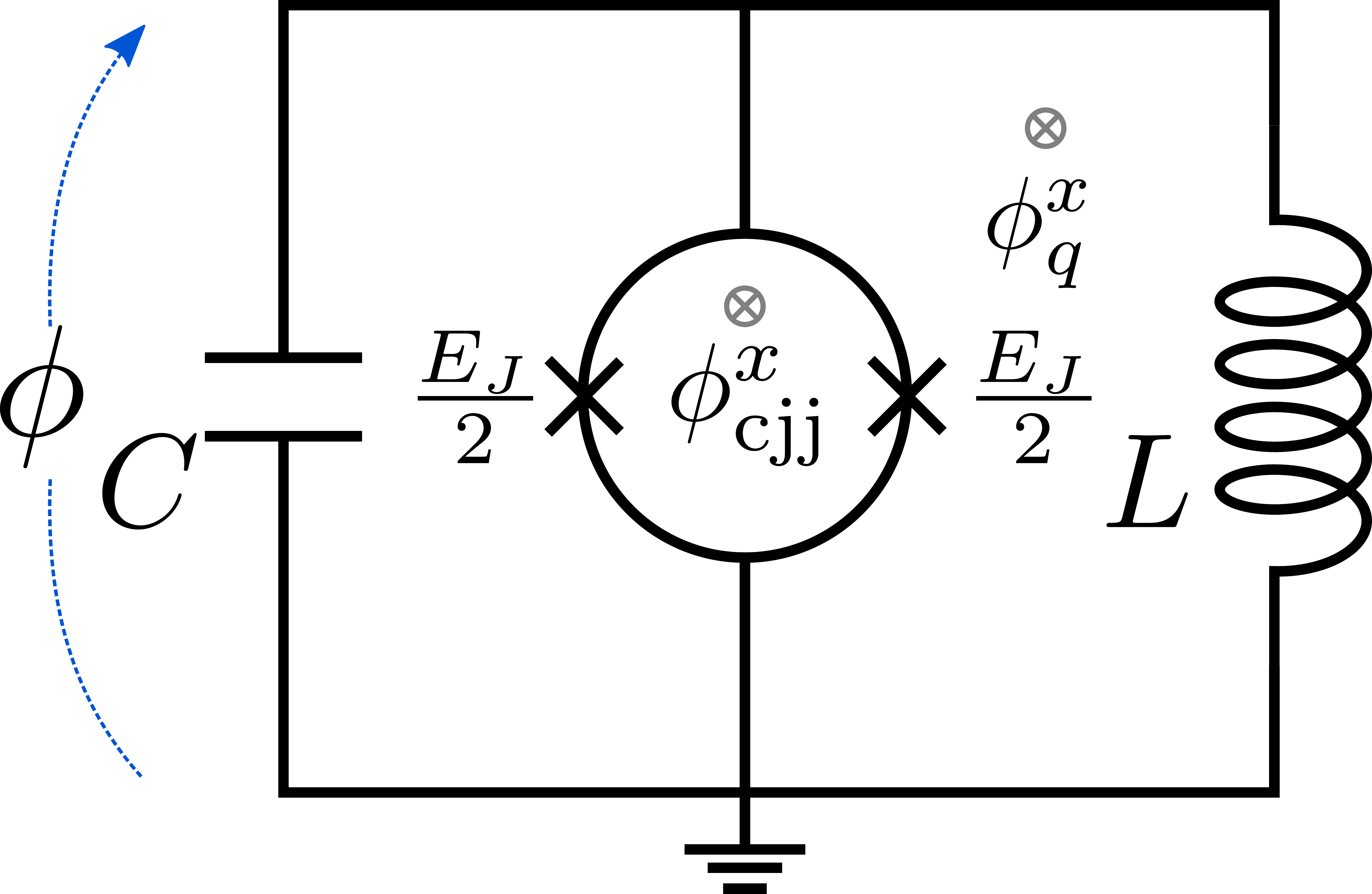}
\caption{Circuit of a flux qubit. The dynamical variable $\phi$ is given by $\phi = \frac{2 \pi \Phi}{\Phi_0}$ with $\Phi$ the flux across the inductor. Analogously, the external fluxes in the superconducting loops $\Phi_{\alpha}^x$, $\alpha \in \{\mathrm{cjj, q}\}$ are given in terms of $\phi_{\alpha}^x = \frac{2 \pi \Phi_{\alpha}^x}{\Phi_0}$.}
\label{fig::fq}
\end{figure}

We briefly review the Hamiltonian of the flux qubit circuit and its mapping to a qubit model. A similar discussion can be found in Refs. \cite{michielsen2020, harris2010}. The basic circuit of a compound Josephson junction rf-SQUID flux qubit is shown in Fig. \ref{fig::fq}. Notice that in this circuit we are neglecting the small inductance of the SQUID loop. While there are also other flux qubit designs \cite{orlando1999, youNori2007, harris2010, Yan2016}, we here focus on this simple circuit since it captures the fundamental physics behind the flux qubit and it is also the design used in Ref.~\cite{ozfidan2020}. 

\par  

The Hamiltonian of the circuit in Fig. \ref{fig::fq} reads
\begin{equation}\label{eq::hfq}
H = 4 E_C \hat{q}^2 + \frac{E_L}{2} \hat{\phi}^2 - E_J^{\mathrm{eff}} \bigl(\phi_{\mathrm{cjj}}^x \bigr)  \cos \bigl (\hat{\phi} + \phi_q^x \bigr),
\end{equation}
where we defined the charging energy $E_C = \frac{e^2}{2 C}$ and the inductive energy $E_L = \frac{\Phi_0^2}{4 \pi^2 L}$. The external flux in the SQUID loop $\phi_{\mathrm{cjj}}^x$ allows to control the effective Josephson energy via the relation
\begin{equation}\label{eq::jeff}
E_J^{\mathrm{eff}} \bigl(\phi_{\mathrm{cjj}}^x \bigr) = E_J \cos \biggl(\frac{\phi_{\mathrm{cjj}}^x }{2} \biggr).
\end{equation}
\begin{figure}
\centering
\begin{subfigure}[h]{0.5 \textwidth}
\centering
\includegraphics[width=0.9 \textwidth]{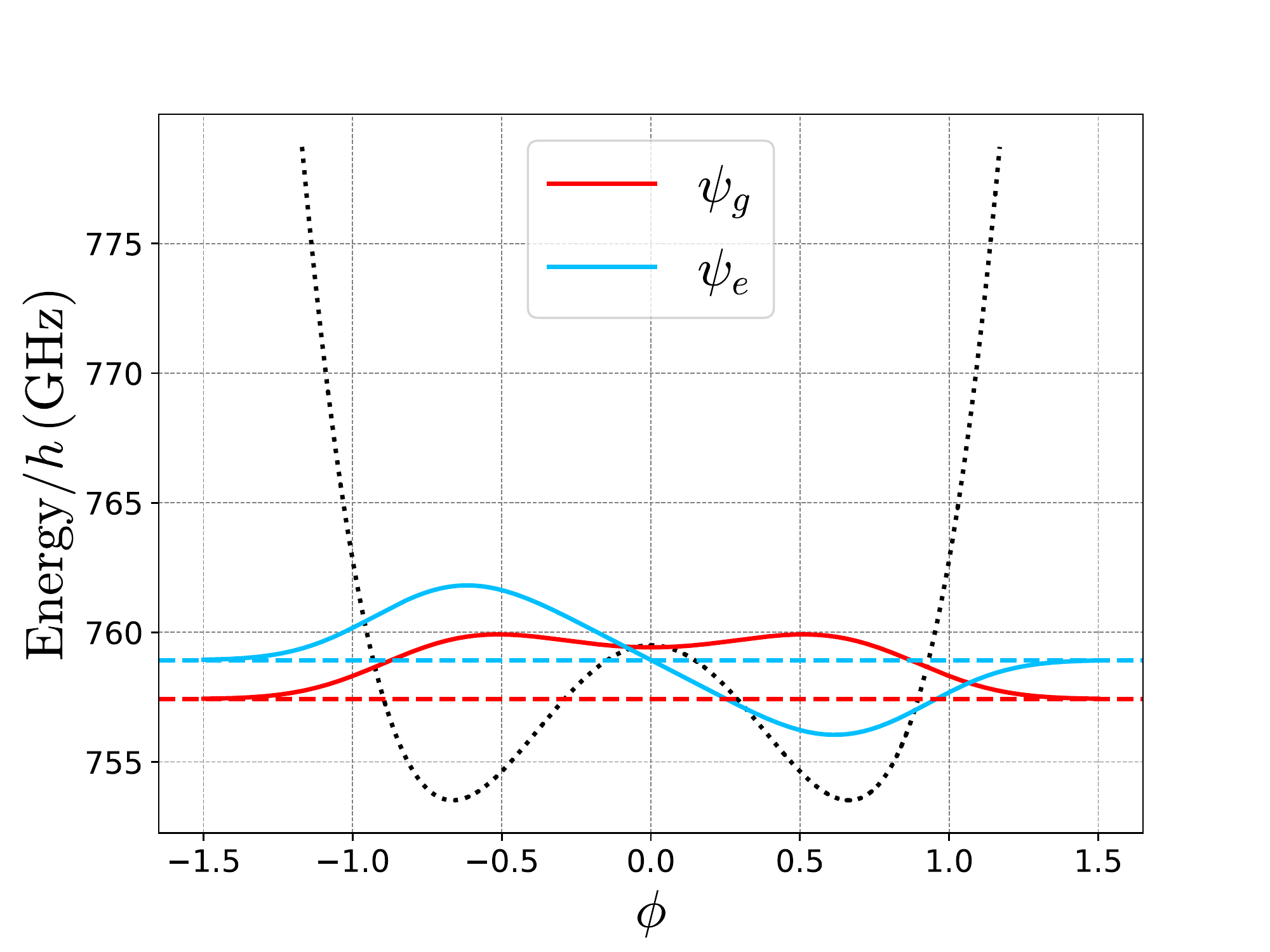}
\subcaption{}
\label{fig::ge}
\end{subfigure}
\begin{subfigure}[h]{0.5 \textwidth}
\centering
\includegraphics[width=0.9\textwidth]{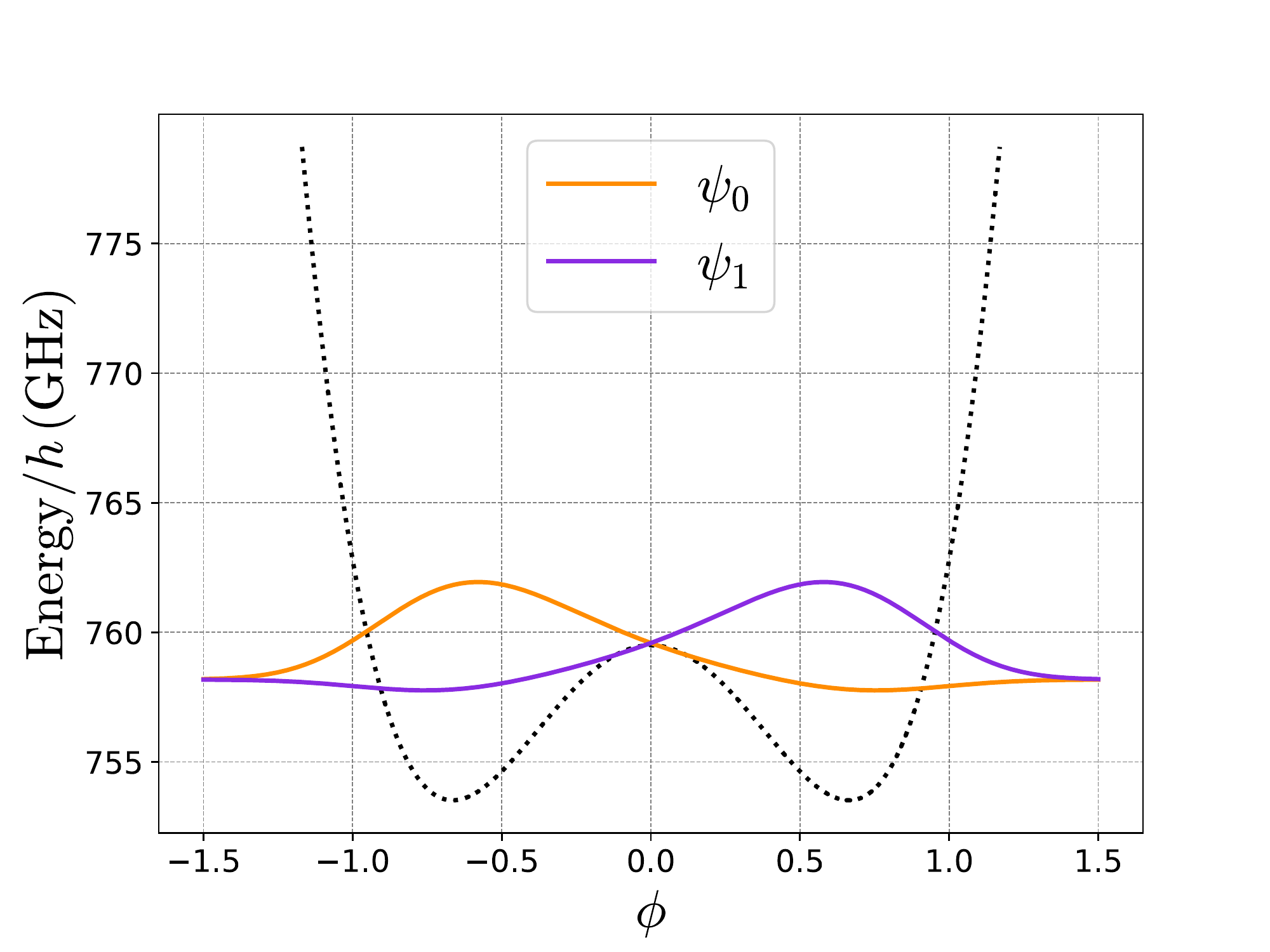}
\subcaption{}
\label{fig::cbasis}
\end{subfigure}
\caption{Flux qubit with symmetric potential. a) Ground and first-excited wave-functions in $\phi$, and their energies as dashed lines.
b) Computational basis states obtained as symmetric and anti-symmetric combination of the first two eigenstates. The relevant parameters are taken as in Table \ref{tab:params}. }
\end{figure}

In what follows we will assume $\phi_{\mathrm{cjj}}^x \in [-\pi, \pi)$. \par 
The flux qubit is operated in the regime $E_L, E_{J}^{\mathrm{eff}} \gg E_C$, $\phi_{q}^x \approx \pi$ and $E_J^{\mathrm{eff}}/E_{L} \gtrsim 1$ \cite{wendinShumeiko, harris2010}. 

With these conditions the potential becomes a double well potential. The computational qubit basis is defined by considering the case of a symmetric potential obtained for  $\phi_{q}^x = \pi$. In this case the eigenstates obey a parity symmetry and consequently they are either even or odd in the flux representation. An example of the wave-functions for the ground-state $\ket{g}$ and first-excited state $\ket{e}$ is shown in Fig. \ref{fig::ge}. The computational basis is defined by taking symmetric and anti-symmetric superpositions of $\ket{g}$ and $\ket{e}$ as 
\begin{subequations}\label{eq:cbasisfq}
\begin{align}
\ket{0} &= \frac{1}{\sqrt{2}}(\ket{g} + \ket{e}) \\
\ket{1} &= \frac{1}{\sqrt{2}}(\ket{g} - \ket{e}).
\end{align}
\end{subequations}

As we see from Fig. \ref{fig::cbasis}, the computational basis states $\ket{0}$, $\ket{1}$ are localized on the left and right well, respectively. They correspond to anti-clockwise and clockwise average persistent currents in the loop formed by the inductor and the SQUID in Fig. \ref{fig::fq}. The energy difference between ground and first-excited state in case of a symmetric potential 
\begin{equation}\label{eq::tcoup}
\Delta = \frac{E_e - E_g}{h}
\end{equation} 
is usually called the tunnel coupling. \par 
By projecting onto the computational subspace the Hamiltonian with symmetric potential is 
\begin{equation}
H_q/h = -\frac{\Delta}{2} X. 
\end{equation}
The external flux in the SQUID loop $\phi_{\mathrm{cjj}}^x$ can be used to control the height of the barrier and thus,  the tunnel coupling $\Delta$. \par 
Let us now consider the asymmetric case in which we slightly bias $\phi_{q}^x$ away from $\pi$, \i.e., we take $\phi_{q}^x = \pi + \delta_q^{x}$. By expanding the cosine in Eq. \eqref{eq::hfq} to first order in $\delta_q^x$ we obtain
\begin{multline}
H \approx 4 E_C \hat{q}^2 + \frac{E_L}{2} \hat{\phi}^2 + E_J^{\mathrm{eff}} \bigl(\phi_{\mathrm{cjj}}^x \bigr)  \cos  \hat{\phi} \\
-E_J^{\mathrm{eff}} \delta_{q}^x \sin \hat{\phi} = H_{\mathrm{sym}} + V.
\end{multline} 
The Hamiltonian is now given by the Hamiltonian in the symmetric case plus a perturbation
\begin{equation}
V = - E_J^{\mathrm{eff}} \delta_{q}^x \sin \hat{\phi}.
\end{equation}
By projecting $V$ onto the computational subspace we obtain a term that is (by design)  diagonal in the computational basis, since $\braket{0| \sin \hat{\phi}|1} = 0$. 
In addition, we can also neglect the coupling that $V$ induces to other energy levels with higher energy. The projected qubit Hamiltonian in the asymmetric case then becomes
\begin{equation}
H_q/h = -\frac{\Delta}{2} X -  \frac{\varepsilon}{2} Z,
\end{equation}
where we defined
\begin{multline}\label{eq::eps}
\varepsilon = \frac{E_J^{\mathrm{eff}}}{h} \delta_{q}^x \bigl(\braket{0 | \sin \hat{\phi} | 0} -  \braket{1 | \sin \hat{\phi} | 1} \bigr) = \\
2 \frac{E_J^{\mathrm{eff}}}{h} \delta_{q}^x \braket{0 | \sin \hat{\phi} | 0}. 
\end{multline}

For relatively large asymmetry of the potential $\delta_{q}^x$ and small $\Delta$ the computational basis states also become the eigenbasis. In addition, the parameter $\delta_{q}^x$ can be used to control the strength of the parameter $\varepsilon$, independently of the tunnel coupling $\Delta$. The independent tunability of $\Delta$ and $\varepsilon$ by means of external fluxes is one of the features that makes flux qubits suited for quantum annealing algorithms \cite{Johnson_2010}. In addition, an inductive coupling between two flux qubits with index $k, l$ would give a term in the Hamiltonian proportional to $\hat{\phi}_k \hat{\phi}_l$, which when projected onto the computational basis gives a term $\sim Z_k Z_l$, thus realizing a TIM.

\subsection{Two-flux qubit Hamiltonian with symmetric double well potentials}\label{app::paritysym}

We analyze a two-flux qubit Hamiltonian with inductive and capacitive couplings, as in Fig. \ref{fig::fqcoup}, and choose symmetric double wells for both flux qubits. As we will see this implies parity symmetry of the Hamiltonian. Based on this symmetry we show that the effective qubit Hamiltonian is always stoquastic by a simple basis change, at any order in SW perturbation theory \footnote{The analysis applies also to fluxonium qubits in a double well configuration since the Hamiltonian is effectively the same as for flux qubits, but in the parameter regime $E_L < E_C < E_J^{\mathrm{eff}}$.}. This shows that asymmetry in the flux qubit potential is necessary to get an effective non-stoquastic qubit Hamiltonian.

First, let the global parity operator $\hat{\pi}$ be a unitary, Hermitian operator, defined via its action on $\hat{\phi}_k$, $\hat{q}_k$ as 
\begin{equation}
\hat{\pi} \hat{\phi}_k \hat{\pi} = - \hat{\phi}_k, \quad \hat{\pi} \hat{q}_k \hat{\pi} = - \hat{q}_k.
\end{equation}
Note that $\hat{\pi}=\Pi_k \exp(i \pi a_k^{\dagger} a_k)$ with $\hat{\phi}_k=\frac{1}{\sqrt{2}}(a_k+a^{\dagger}_k), \hat{q}_k=\frac{i}{\sqrt{2}}(a^{\dagger}_k-a_k)$.

Since $\hat{\pi}^2 = 1$ the parity operator has eigenvalues $\pm 1$. We call an operator $O$ parity symmetric when $\hat{\pi} O \hat{\pi}=O$.

The Hamiltonian of the two flux qubits in case $\phi_{q}^x = \pi$ (symmetric double wells) reads
\begin{multline}\label{eq::h2q}
H = \sum_{k=1}^2 4 E_{C k} \hat{q}_k^2 + \frac{1}{2} E_{Lk} \hat{\phi}_k^2 + E_{Jk}^{\mathrm{eff}}  \cos \hat{\phi}_k \\+ 8 E_{C12} \hat{q}_1 \hat{q}_2 + E_{L12} \hat{\phi}_1 \hat{\phi}_2,
\end{multline}       
and we can define the single flux qubit Hamiltonian
\begin{equation}\label{eq::hfq_sym}
H_{k} = 4 E_{Ck} \hat{q}_k^2 + \frac{1}{2} E_{Lk} \hat{\phi}_k^2 + E_{Jk}^{\mathrm{eff}}  \cos \hat{\phi}_k,
\end{equation}
$k=1, 2$, with $E_{Ck}$ and $E_{Lk}$ the diagonal elements of the charging energy and inductive energy matrix respectively. 

We can write Eq. \eqref{eq::h2q} as
\begin{equation}
H = H_0 + V,
\end{equation}
with $H_0=H_1+H_2$, \i.e., the uncoupled flux qubit Hamiltonians, and 
\begin{equation}\label{eq::h_intp}
V = 8 E_{C12} \hat{q}_1 \hat{q}_2 + E_{L12} \hat{\phi}_1 \hat{\phi}_2,
\end{equation}
where $E_{C12}$ and $E_{L12}$ denote the off-diagonal element of $\bm{E_C}$ and $\bm{E_L}$ respectively.  Clearly, both $H_0$ as well as $V$ are invariant under the global parity transformation $\hat{\pi}$.
 
 In the remaining part of this section the index $k$ will always be $k=1,2$. The Hamiltonians $H_{k}$ admit only bound states as eigenstates. As a consequence  the average of $\hat{q}_{k}$ in any eigenstate $\ket{\psi}_{k}$ of $H_{k}$ is zero, \i.e., $\braket{\psi| \hat{q}_{k} | \psi}_{k} = 0$ \cite{messiah}.

Due to parity symmetry, the eigenstates of Hamiltonians $H_0$ and $H$ can be chosen as eigenstates of the parity operator with eigenvalues $\pm 1$. This implies that the eigenstate wave-functions in flux are either even or odd functions, and that
$ \braket{\psi| \hat{\phi}_k | \psi }_k= 0$.

We now assume the validity of first order perturbation theory in the eigenbasis of the $H_k$ and obtain an effective two-qubit Hamiltonian by projecting Eq. \eqref{eq::h2q} onto the subspace $\mathcal{P}_0$ spanned by the first two levels of each subsystem $\ket{g}_{k}$, $\ket{e}_{k}$. This consists in applying the projector
\begin{equation*}
P_0 = P_1 \otimes P_2 = \bigl(\ket{g}\bra{g}_1 + \ket{e}\bra{e}_1 \bigr) \otimes \bigl(\ket{g}\bra{g}_2 + \ket{e}\bra{e}_2 \bigr)
\end{equation*}
to the Hamiltonian in Eq. \eqref{eq::h2q}. By defining our Pauli operators in the eigenbasis as
\begin{subequations}\label{eq:paulis}
\begin{align}
X_k = \ket{g}\bra{e}_k + \ket{e}\bra{g}_k,
\end{align}
\begin{align}
Y_k = -i \ket{g}\bra{e}_k + i \ket{e}\bra{g}_k,
\end{align}
\begin{align}
Z_k =  \ket{g}\bra{g}_k - \ket{e}\bra{e}_k,
\end{align}
\end{subequations}
the effective qubit Hamiltonian reads
\begin{multline}\label{eq:h2q_ic}
H_{2 q}/h = P_0 H P_0 = -\frac{\Delta_1}{2} Z_1 -\frac{\Delta_2}{2} Z_2 + \\
J_{XX} X_1 X_2 + J_{YY} Y_1 Y_2,
\end{multline}
where the tunnel couplings $\Delta_{1,2}$ are defined as in Eq. \eqref{eq::tcoup}, $J_{YY}$ is given by 
\begin{equation}
J_{YY} = -\frac{E_{C12}}{h} \braket{g| \hat{q}_1 | e}_1 \braket{g| \hat{q}_2 | e}_2
\end{equation} 
and the XX coupling $J_{XX}$ is
\begin{equation}
J_{XX} =  \frac{E_{L12}}{h} \braket{g| \hat{\phi}_1 | e}_1 \braket{g| \hat{\phi}_2 | e}_2.
\end{equation}

The Hamiltonian in Eq. \eqref{eq:h2q_ic} can always be made stoquastic by simple Clifford transformations. The conditions for stoquasticity is here that $J_{XX} \leq -|J_{YY}|$.

If $\lvert J_{XX} \rvert \ge \lvert J_{YY} \rvert$ apply the transformation $X_1 \mapsto -\mathrm{sign}(J_{XX}) X_1, Z_1 \mapsto -\mathrm{sign}(J_{XX}) Z_1$ and then the Hamiltonian is stoquastic. If $\lvert J_{XX} \rvert < \lvert J_{YY} \rvert$ apply the transformation that exchanges $X$ and $Y$ on both qubits, and use the previous transformation.\par 
The previous result relies on the validity of the projection onto the computational subspace. A natural question to ask is whether the effective qubit Hamiltonian can still always be made stoquastic if the perturbation theory is refined. We here show this is indeed the case by using a Schrieffer-Wolff transformation \cite{bravyiDiVincenzo}. Note that this was also used in Ref. \cite{Consani_2020} to obtain the effective Hamiltonian of two capacitively and inductively coupled flux qubits, and to study its stoquasticity. \par 

To properly discuss the SW transformation we recall some notions from Ref. \cite{bravyiDiVincenzo}. Let $\lambda_{\mathcal{P}_0}^{\mathrm{min}}$ and $\lambda_{\mathcal{P}_0}^{\mathrm{max}}$ be the minimum and maximum eigenvalues of $H_0$ with eigenvectors in $\mathcal{P}_0$, respectively, and let $\mathcal{I}_0 = [\lambda_{\mathcal{P}_0}^{\mathrm{min}}, \lambda_{\mathcal{P}_0}^{\mathrm{max}}] \subseteq \mathbb{R}$. We define the energy gap $\Lambda = \lambda_{\mathcal{Q}_0}^{\mathrm{min}} - \lambda_{\mathcal{P}_0}^{\mathrm{max}}$, where $\lambda_{\mathcal{Q}_0}^{\mathrm{min}}$ is the minimum eigenvalue of $H_0$ whose eigenvectors is in the complement subspace $\mathcal{Q}_0$ of $\mathcal{P}_0$. We introduce a new interval $\mathcal{I} = [\lambda_{\mathcal{P}_0}^{\mathrm{min}}-\Lambda/2, \lambda_{\mathcal{P}_0}^{\mathrm{max}} + \Lambda/2] \subseteq \mathbb{R}$, and the subspace $\mathcal{P}$ with projector $P$ spanned by the eigenvectors of $H$ with eigenvelue in $\mathcal{I}$. Our general goal is to obtain an effective Hamiltonian that is block-diagonal with respect to $P_0$ and $Q_0$, \i.e., $P_0 H_{\mathrm{eff}} Q_0 = Q_0 H_{\mathrm{eff}} P_0=0$ and has the same spectrum as $H$. In particular, by projecting the effective Hamiltonian onto $P_0$ we obtain a reduced Hamiltonian $H_{0 \mathrm{eff}} = P_0 H_{\mathrm{eff}} P_0$ with spectrum in $\mathcal{I}$. The SW transformation is a unitary transformation defined as
\begin{equation}\label{eq:sw}
U = \exp (S) = \sqrt{(1 - 2 P_0) (1 - 2 P)},
\end{equation}
where $S$ is a block-off-diagonal, anti-hermitian operator with respect to $P_0, Q_0$. In order for the SW to be uniquely defined we require the condition $\left\lVert S \right\rVert < \pi/2$ \cite{bravyiDiVincenzo}. The (exact) effective Hamiltonian $H_{0 \mathrm{eff}}$ is
\begin{equation}
H_{0 \mathrm{eff}} = P_0 U H U^{\dagger} P_0. 
\end{equation}
Since the Hamiltonians $H$ and $H_0$ are invariant under parity transformations, also the projectors $P$, $P_0$ satisfy the parity symmetry. Consequently also the unitary $U$ in Eq. \eqref{eq:sw} and the generator $S$ are parity symmetric, and thus also $H_{0 \mathrm{eff}}$, \i.e.,
\begin{equation}
\hat{\pi} H_{0 \mathrm{eff}} \hat{\pi} = H_{0 \mathrm{eff}}. 
\end{equation}
Since $\hat{\pi} \ket{g}_k = +1 \ket{g}_k$ and  $\hat{\pi} \ket{e}_k = -1 \ket{e}_k$, the parity operator $\hat{\pi}$ acts on the Pauli operators defined in Eq. \eqref{eq:paulis} as
\begin{equation}
\hat{\pi} X_k \hat{\pi} = -X_k, \quad \hat{\pi} Y_k \hat{\pi} = -Y_k, \quad \hat{\pi} Z_k \hat{\pi} = Z_k.
\end{equation}

Thus, the only terms allowed in $H_{0 \mathrm{eff}}$ in order to satisfy the parity symmetry, and the fact that the Hamiltonian is real, are local $Z_{1, 2}$ and the interactions $X_1 X_2, Y_1 Y_2, Z_1 Z_2$. 

Hence, compared to the lowest-order SW projection only the $Z_1 Z_2$ term can be added, and since this term is diagonal we can employ the same Clifford transformations as before to cure the non-stoquasticity. 

We note that this observation does not immediately generalize to multiple flux qubits as the SW transformation may introduce $k$-local terms and it is not clear whether the parity symmetry would suffice in that case.


\bibliography{biblioStoqSCQ}

\end{document}